\documentclass[usenatbib]{mn2e}
\usepackage{natbib}
\usepackage{apjfonts}
\usepackage{amsmath}

\usepackage{epsfig}
\usepackage{graphicx}
\usepackage{amssymb}
\usepackage{aas_macros}
\usepackage{color}

\newcommand{\pc}{\,{\rm pc}}
\newcommand{\kpc}{\,{\rm kpc}}
\newcommand{\kev}{\,{\rm keV}}
\newcommand{\ev}{\,{\rm eV}}

\newcommand{\oversim}[2]{\protect{\mbox{\lower0.5ex\vbox{%
   \baselineskip=0pt\lineskip=0.2ex
   \ialign{$\mathsurround=0pt #1\hfil##\hfil$\crcr#2\crcr\sim\crcr}}}}} 
\newcommand{\bb}[1]{\ifmmode \mbox{\boldmath $ #1$} \else  \mbox{\boldmath $#1$} \fi}
\catcode`\"=\active\let"=\"  

\def\3{{\ss} }

\def\c12{{1\over 2}}

\def\d{{\rm d}}   
   
\def\plusplus{\raise 0.3ex\hbox{${\scriptstyle ++}$}{}}

\def\and{{{\rm M}31}}

\bibliography{biblio}

\begin{document}   
\title[Stochastic substructure forces]{Fluctuations of the gravitational field generated by a random population of extended substructures}

\author[Jorge Pe\~{n}arrubia]{Jorge Pe\~{n}arrubia$^{1}$\thanks{jorpega@roe.ac.uk}\\
$^1$Institute for Astronomy, University of Edinburgh, Royal Observatory, Blackford Hill, Edinburgh EH9 3HJ, UK\\
}
\maketitle  

\begin{abstract}
A large population of extended substructures generates a stochastic gravitational field that is fully specified by the function $p({\bf F})$, which defines the probability that a tracer particle experiences a force ${\bf F}$ within the interval ${\bf F},{\bf F}+\d{\bf F}$.
This paper presents a statistical technique for deriving the spectrum of random fluctuations directly from the number density of substructures with known mass and size functions.
Application to the subhalo population found in cold dark matter simulations of Milky Way-sized haloes shows that, while the combined force distribution is governed by the most massive satellites, the fluctuations of the {\it tidal} field are completely dominated by the smallest and most abundant subhaloes.
In light of this result we discuss observational experiments that may be sufficiently sensitive to Galactic tidal fluctuations to probe the ``dark'' low-end of the subhalo mass function and constrain the particle mass of warm and ultra-light axion dark matter models.
\end{abstract}   

\begin{keywords}
Galaxy: kinematics and dynamics; galaxies: evolution. 
\end{keywords}

\section{Introduction}\label{sec:intro}

Observations of the cosmic microwave background provide robust evidence for the existence of large amounts of non-baryonic matter (e.g Planck collaboration 2014) that behaves like a perfect fluid on large scales (e.g. Peacock 1999). In theory, deviations from the perfect-fluid model are expected to arise on scales comparable to the free-streaming length of the DM particle candidates. Below this scale, fluctuations of the power spectrum are heavily suppressed, de facto imposing a truncation at the low end of the halo mass function (e.g. Benson 2017 and references therein). However, for most DM-particle candidates the truncation falls well below the threshold of galaxy formation (White \& Rees 1978; Bullock et al. 2000), which greatly complicates observational tests. For example, in cold dark matter (CDM) models made of WIMPs with masses $\sim 1$~GeV (e.g. Bertone et al. 2005) the cut-off is placed on planet-size mass scales, $M_1\sim 10^{-6}M_\odot$ (e.g. Schmid et al. 1999; Hofmann et al. 2001; Green et al. 2005; Loeb \& Zaldarriaga 2005; Diemand et al. 2005), which leads to one of the most striking predictions from the cold dark matter (CDM) paradigm, namely the existence of a very large number of self-gravitating ``microhaloes'' with subsolar masses devoid of baryons (i.e. `dark').

The lack of visible matter in these objects together with their tiny masses makes the detection of DM microhaloes extremely challenging. Current observational efforts to test the presence of dark subhaloes in the Milky Way range from searching for gamma-ray annihilation signals (e.g. Ackermann et al. 2014; Bringmann et al. 2014) to detecting gaps in narrow tidal streams induced by close encounters with individual subhaloes (Ibata et al. 2002; Johnston et al. 2002; Yoon et al. 2011; Carlberg 2013; Erkal \& Belokurov 2015; Ngan et al. 2016, Erkal et al. 2016; Bovy et al. 2017), with no unambiguous results to date.
Strongly-lensed galaxies provide complementary constraints on the clumpiness of dark matter haloes in the inner-most region of galaxies (Koopmans 2005; Vegetti \& Koopmans 2009; Li et al. 2013; Vegetti et al. 2014). 

In general, current observational tests tend to probe the high-mass end of the subhalo mass function. For example, the sizes of stream gaps produced by encounters with DM substructure in Milky Way-like galaxies are too small to be detectable for subhalo masses $M\lesssim 10^5$--$10^6M_\odot$ (Carlberg 2012; Erkal\& Belokurov 2015; Erkal et al. 2016; Bovy et al. 2017), with the added complexity that tidal heating by giant molecular clouds (GMCs) becomes non-negligible on similar mass scales, $M_{\rm GMC}\lesssim 10^7M_\odot$ (Amorisco et al. 2016). Perturbations of Einstein rings around lensed galaxies are also expected to be dominated by relatively massive subhaloes with $M\gtrsim 10^7M_\odot$ (Li et al. 2016).

To the observational challenge of finding tiny invisible objects one must add the extreme dynamical range of CDM subhaloes.
Collision-less $N$-body cosmological simulations that explore the properties of CDM substructures within Milky Way-size systems, e.g. the Aquarius Project (Springel et al. 2008), Via Lactea (Diemand et al. 2007) and GHALO (Stadel et al. 2009), have particle-mass resolution $10^4$--$10^5M_\odot$, which lies $\sim 10$ orders of magnitude above the mass scale of microhaloes. 

To date 
it remains unclear to what extent the dearth of unresolved minihaloes may affect the predictions on the gamma-ray annihilation signals (e.g. Koushiappas 2009), and/or the dynamics of self-gravitating systems orbiting in the host galaxy.
Similarly, the survivability of microhaloes in the tidal field of the parent galaxy is still poorly understood. On the one hand, fly-by encounters with individual stars (Green \& Goodwin 2007) and the tidal field of the Milky Way disc (D'Onghia et al. 2010; Errani et al. 2017) may wipe out a large fraction of the subhalo population with small orbital pericentres.
On the other hand, the steep density profile of microhaloes (Anderhalden \& Diemand 2013; Angulo et al. 2016) greatly increases their resilience to tidal mass stripping (Goerdt et al. 2007; Pe\~narrubia et al. 2010).

For those interested in the history of science the search for dark matter microhaloes bears unmistakable resemblance to the discussions on the 'reality of molecules' during the 19th century. The existence of astronomical numbers of {\it invisible molecules} had been theoretically postulated by Avogrado in 1811 in order to explain the observed thermodynamical properties of gases. A heated epistemologic debate on whether molecules were real physical entities or just a useful mathematical construct lasted for nearly a century until Einstein (1905) introduced a revolutionary method for inferring the number and size of these objects from observations of the Brownian motion of particles suspended on the surface of a liquid. Given the very large number of degrees of freedom involved in the problem, Einstein entirely abandoned the classical Newtonian approach of solving the phase-space trajectories of individual molecules from deterministic equations of motion, focusing instead on a statistical description of the response of macroscopic objects to repeated interactions with a very large number of small particles. 

This contribution follows a similar line of reasoning. Rather than following the dynamical evolution and the orbits of individual DM microhaloes in order to simulate their gravitational interactions with tracer particles, our goal is to construct a {\it probability} function that fully specifies the combined gravitational field generated by subhalo ensembles. As an illustration, in Section~\ref{sec:fluctu} we run numerical experiments that illustrate the stochastic nature of the combined gravitational force generated by a large subhalo population. Following up the work of Holtsmark (1919) we develop a simple statistical technique for deriving the spectrum of force fluctuations directly from the number density and the mass \& size functions of extended substructures. In Section~\ref{sec:tides} we extend the analysis to the combined tidal field generated by these objects. Interestingly, we find that the spectrum of tidal forces induced by DM subhaloes is particularly sensitive to the low-mass truncation of the halo mass function, suggesting that the existence of `dark' satellites could in principle be tested with observational experiments that measure fluctuations in the Galactic tidal field. Section~\ref{sec:dis} compares the force and tidal distributions expected in Milky Way-like galaxies for DM haloes made of cold/warm particles and ultra-light axions. Section~\ref{sec:sum} summarizes our finding and discusses further applications of our method.

\section{Stochastic fluctuations of the field}\label{sec:fluctu}
Let us assume that the gravitational acceleration experienced by a tracer particle at a distance ${\bf R}$ from the galaxy centre can be expressed as
\begin{eqnarray}\label{eq:eqmot}
\frac{{\d^2 \bf R}}{\d t^2}=-\nabla\Phi_g({\bf R}) + \sum_{i=1}^{N} {\bf f}_i({\bf R}),
\end{eqnarray}
where $\Phi_g$ is the mean-field gravitational potential of the galaxy, and
\begin{eqnarray}\label{eq:F}
{\bf F}\equiv \sum_{i=1}^{N}{\bf f}_i,
\end{eqnarray}
is the specific force induced by a set of self-gravitating {\it substructures} orbiting in the host galaxy. In systems where $N\gg 1$ we expect ${\bf F}$ to fluctuate stochastically along the phase-space trajectory of the particle, ${\bf R}(t)$.

Solving~(\ref{eq:eqmot}) for an arbitrarily-large population of substructures appears an impossibly difficult task given that the trajectory of the tracer particle is coupled with the equations of motion of each individual substructure.
To attack this problem we shall follow a statistical method originally introduced by Holtsmark (1919) to describe the motion of charged particles in a plasma. 

The crux of our analysis lies in the function $p({\bf F})$, which defines the probability density that the test particle experiences a force in the interval ${\bf F}, {\bf F}+\d{\bf F}$. The simplest derivation assumes that the $N$-substructures are homogeneously distributed within a volume $V=4\pi d^3/3$ around the test particle, such that
\begin{eqnarray}\label{eq:pF}
 p({\bf F})=\frac{1}{V}\int \d^3r_1 \times ...\times \frac{1}{V}\int \d^3r_N\delta\bigg({\bf F}-\sum_i{\bf f}_i\bigg),
\end{eqnarray}
where $\delta$ is the Dirac's delta function. Fourier transforming $p({\bf F})$ yields
\begin{eqnarray}\label{eq:FTpF}
  \tilde p({\bf k})&=&\int \d^3 F e^{i{\bf k}\cdot {\bf F}}p({\bf F})\\ \nonumber 
  &=&\frac{1}{V}\int \d^3r_1 \times ...\times \frac{1}{V}\int \d^3r_N \int \d^3 F e^{i{\bf k}\cdot{\bf F}}\delta\bigg({\bf F}-\sum_j{\bf f}_j\bigg)\\ \nonumber
  &=&\frac{1}{V}\int \d^3r_1 \times ...\times \frac{1}{V}\int \d^3r_N e^{i{\bf k}\sum_j{\bf f}_j}\\ \nonumber
  &=&\bigg[\frac{1}{V}\int \d^3r e^{i{\bf k}\cdot\bf f}\bigg]^N.
\end{eqnarray}
Note that the last equality in Equation~(\ref{eq:FTpF}) implicitly assumes that the $N$-particles are randomly distributed within the volume $V$ or, equivalently, that the forces ${\bf f}_i$ are spatially uncorrelated. The last integral can be re-written as
$$\frac{1}{V}\int \d^3r e^{i{\bf k}\cdot\bf f}=\frac{1}{V}\int_V \d^3r\bigg[ 1-\big(1-e^{i{\bf k}\cdot\bf f}\big)\bigg]=1-\frac{1}{V}\int_V \d^3r\big(1-e^{i{\bf k}\cdot\bf f}\big),$$
which elevated to the $N$-th power becomes
$$\bigg[1-\frac{1}{V}\int_V \d^3r\big(1-e^{i{\bf k}\cdot\bf f}\big)\bigg]^N\approx \exp\bigg[-n\int_V \d^3r\big(1-e^{i{\bf k}\cdot\bf f}\big)\bigg] ~~~~{\rm for} ~~N\gg 1,$$
where $n\equiv N/V$ is the number density of substructures. It is useful to define the function
\begin{eqnarray}\label{eq:phi}
  \phi({\bf k})\equiv n\int_V \d^3r\big(1-e^{i{\bf k}\cdot\bf f}\big),
\end{eqnarray}
such that the inverse Fourier transform of~(\ref{eq:FTpF}) becomes
\begin{eqnarray}\label{eq:pF2}
  p({\bf F})=\frac{1}{(2\pi)^3}\int \d^3k \exp\big[-i{\bf k}\cdot{\bf F}-\phi({\bf k})\big].
\end{eqnarray}
The above derivation can be easily generalized to inhomogeneous distributions of substructures by taking into account that the number density varies with radius, Equation~(\ref{eq:phi}) then becomes (Chandrasekhar 1941; Kandrup 1980; Chavanis 2009)
\begin{eqnarray}\label{eq:phi_inhomog}
  \phi_{\rm in}({\bf k})= \int\d^3r\big(1-e^{i{\bf k}\cdot\bf f}\big) n({\bf r}),
\end{eqnarray}
where ${\bf r}$ is centred at the location of the test particle, and $n({\bf r})$ is the number density profile. On scales $d\lesssim |\nabla n/n|^{-1}$ the number density can be assumed to be roughly constant, $n({\bf R}+{\bf r})\approx n({\bf R})=n$, such that $\phi_{\rm in}({\bf k})\approx \phi({\bf k})$, which is typically known as the {\it local approximation} (see Appendix A).

The function $\phi({\bf k})$ contains all physical information on the number, distribution and masses of the substructure population. Unfortunately, it can be rarely expressed in an analytical form. Below we inspect a few notable exceptions with broad applications in Astronomy. 

\subsection{Point-mass particles}\label{sec:pm}
As a first step it is useful to review the analysis of fluctuations in the gravitational field induced by a random distribution of equal-mass particles (see Kandrup 1980 for a formal description of this method).

The gravitational force exerted by a point-mass particle can be written as
\begin{eqnarray}\label{eq:fpm}
{\bf f}=-\frac{GM }{r^2}\hat{\bf r},
\end{eqnarray}
where $\hat{\bf r}$ is a unit vector. For simplicity, particles are assumed to be homogeneously distributed across the galaxy. Inserting Equation~(\ref{eq:fpm}) into~(\ref{eq:phi}) and writing ${\bf k}\cdot {\bf f}= k f \cos\theta$ yields
\begin{eqnarray}\label{eq:phipm}
  \phi({\bf k})&=& 2\pi n\int_0^{\infty} \d r r^2 \int_{-1}^{+1}\d(\cos \theta)\big(1-\exp[-i k GM\cos \theta/r^2]\big) \\ \nonumber
  &=& 4\pi n \int_0^{\infty} \d r r^2 \big(1-\frac{\sin[k GM /r^2]}{k GM /r^2}\big)\\ \nonumber
  &=& \frac{4}{15}(2\pi GM )^{3/2}n k^{3/2} \equiv ak^{3/2},
\end{eqnarray}
where $a\equiv \frac{4}{15}(2\pi GM )^{3/2}n$.
Note that the resulting function is isotropically oriented in Fourier space, i.e. $\phi({\bf k})=\phi(k)$, which implies that the force distribution must also be isotropic, i.e. $p({\bf F})=p(F)$. Indeed, 
combination of Equations~(\ref{eq:pF2}) and~(\ref{eq:phipm}) leads to the well-known Holtsmark (1919) distribution
\begin{eqnarray}\label{eq:pF_pm}
  p({\bf F})&=&\frac{1}{(2\pi)^2}\int_0^\infty \d k\, k^2\exp(-a k^{3/2})\int_{-1}^{+1}\d x\exp[-i k F x]\\ \nonumber 
  &=&\frac{1}{2\pi^2}\int_0^\infty \d k\, k^2 \exp(-a k^{3/2})\frac{\sin(k F)}{kF}.
\end{eqnarray}

The asymptotic behaviour of this function is studied in Chandrasekhar (1943) and reproduced here for completeness. In the {\it weak-force limit} we can approximate $\sin(k F)\approx kF$, such that
\begin{eqnarray}\label{eq:pF_pm_zero}
  \lim_{F\to 0} p({\bf F})= \frac{1}{2\pi^2}\int_0^\infty \d k\, k^2\exp(-a k^{3/2})=\frac{1}{3 \pi^2 a^2}.
  \end{eqnarray}
  We thus find that the effect of having an increasing number of particles at large distances combines with the declining Keplerian force such that in the weak-force limit $p({\bf F})$ becomes flat.
 Deriving the {\it strong-force limit} requires a number of non-trivial steps. Chandrasekhar (1943) finds
\begin{eqnarray}\label{eq:pF_pm_inf}
  \lim_{F\to \infty} p({\bf F})\approx \frac{15}{8}\frac{a}{(2\pi)^{3/2}}F^{-9/2}=\frac{1}{2}(GM)^{3/2}n F^{-9/2}.
  \end{eqnarray}
It is important to stress that the large-force behaviour is entirely dominated by the contribution of the nearest particle. This can be straightforwardly shown by writing the probability of finding the closest particle within the volume ${\bf r},{\bf r}+\d^3r$ as (see Appendix A of Chavanis 2009)
\begin{eqnarray}\label{eq:pr_closest}
  p({\bf r})\d^3 r\sim \exp\big(-\frac{4}{3}\pi r^3 n\big )4\pi r^2 n \d r,
    \end{eqnarray}
which peaks at $D\equiv (2\pi n)^{-1/3}$, and making the transformation $p({\bf r}) \d^3 r=p({\bf F})\d^3 F$, where $|{\bf F}|=G M/r^2$ is the force induced by the nearest particle. For nearby objects, $r^3 n\ll 1$, Equation~(\ref{eq:pr_closest}) reduces to $p({\bf r})\d^3 r\approx 4\pi r^2 n \d r$, whereas $\d r=1/2 (GM)^{1/2}F^{3/2}d F$. Hence,
\begin{eqnarray}\label{eq:p0}
  p_0({\bf F})=n\frac{r^2\d r}{F^2\d F}=\frac{1}{2}(G M)^{3/2}n F^{-9/2},
   \end{eqnarray}
  as in Equation~(\ref{eq:pF_pm_inf}).

It is useful to illustrate the above results by means of two simple numerical experiments. In experiment A we randomly generate independent samples of $N$ equal-mass particles distributed homogeneously within a volume $V=4\pi d^3/3$. For each particle ensemble we measure the total force ${\bf F}=\sum_{i=1}^N {\bf f}_i=-\sum_{i=1}^N GM{\bf r}/r^3$. In experiment B we assign isotropic velocities to one random set of point-masses generated in experiment A,
such that the resulting orbital distribution is in equilibrium within a a harmonic spherical potential $\Phi_g(r)=\frac{1}{2}\Omega_0 r^2$. The orbits of individual particles are integrated in the smooth potential for 10 dynamical times, where $t_{\rm dyn}=(3\pi /32)^{1/2}\Omega_0^{-1}$. Clearly, experiment A is designed to mimic the mathematical framework devised in Section~\ref{sec:fluctu}, whereas experiment B provides a simplistic representation of the dynamics of point-mass particles in a galaxy.

\begin{figure}
\begin{center}
\includegraphics[width=85mm]{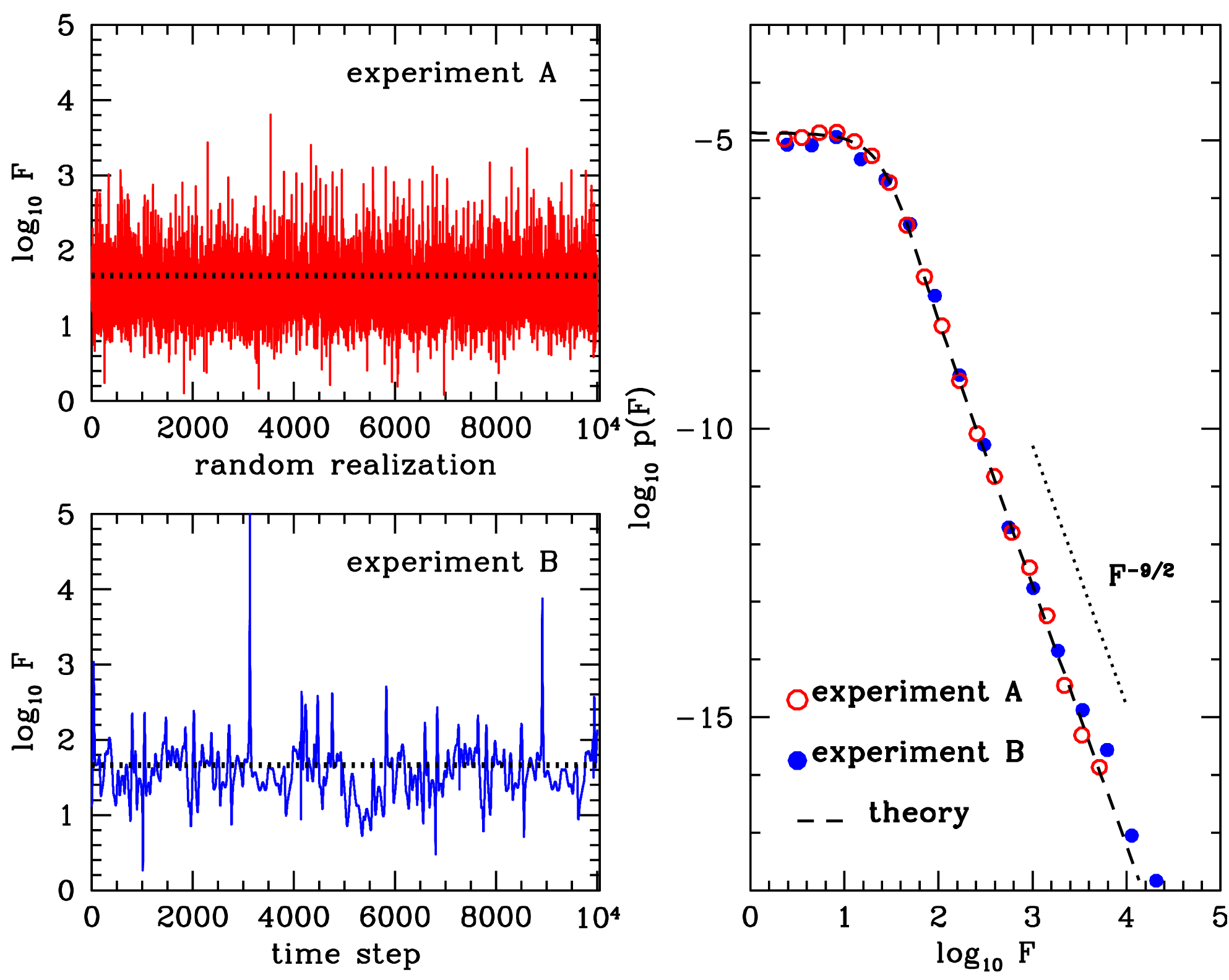}
\end{center}
\caption{Total force ${\bf F}=-\sum_{i=1}^N GM{\bf r}/r^3$ induced by $N=5\times 10^4$ particles with a mass $M=G=1$ homogeneously distributed over a sphere of radius $d=10$. {\it Upper-left panel} shows the force measured for $10^4$ random ensembles of equal-mass particles. {\it Lower-left panel shows} the time-varying force experienced by a single tracer particle ensemble orbiting in a harmonic potential $\Phi_g=\Omega_0 r^2/2$, and $\Omega_0=1$. The orbits of individual particles are integrated for 10 dynamical times, where $t_{\rm dyn}=(3\pi /32)^{1/2}\Omega_0^{-1}$. In both panels the black-dotted line marks the mean force, $\langle F\rangle$. {\it Right panels} shows the probability density $p({\bf F})$ measured from both experiments. The theoretical expectation (long-dashed line) is derived from Equation~(\ref{eq:pF_pm}). }
\label{fig:evolgen}
\end{figure}

Fig.~\ref{fig:evolgen} shows 
the combined force induced by $N=5\times 10^4$ equal-mass particles with $G=M=1$ homogeneously distributed over a sphere of radius $d=10$.
Comparison between the upper- and lower-left panels shows clear differences between the force measured from random statistical samples and that generated by an equilibrium ensemble at different time-steps. The reason behind the mismatch can be traced to the different {\it speed of fluctuations}, $\d F/\d t$, in experiments A and B. In dynamical-equilibrium systems such as experiment B this quantity is modulated by time-scale of the orbital motions (see Chandrasekhar \& von Neumann 1942 for a detailed discussion), whereas in experiment A the speed of fluctuations is random by construction.
Yet, the right panel shows that the {\it probability} of measuring a force in the interval ${\bf F},{\bf F}+\d{\bf F}$ are statistically indistinguishable. Equation~(\ref{eq:pF_pm}) (long-dashed line) matches the numerical distributions, in agreement with the early $N$-body results of Ahmad \& Cohen (1973) and del Popolo (1996). This is an important result, as it shows that the probability density $p({\bf F})$ fully specifies the stochastic perturbations induced by an ensemble of equal-mass particles moving in an equilibrium orbital configuration in the host potential.

The specific force induced by a random distribution of particles upon a test star fluctuates around a mean value $\langle F\rangle$, which for ease of reference is marked with horizontal dotted-lines in the left panels of Fig.~\ref{fig:evolgen}. The mean force can be calculated analytically from Equation~(\ref{eq:pF_pm}) by changing the integration variable to $z=k F$ and integrating over the volume $\d^3 F=4\pi F^2\d F$, which yields
\begin{eqnarray}\label{eq:F1}
  \langle F\rangle &=&\int_0^\infty\d^3 F p({\bf F})F \\ \nonumber
  &=&\frac{2}{\pi}\int_0^\infty \d z \sin(z)z\int_0^\infty\d F \exp\big(-az^{3/2}F^{-3/2}\big)\\ \nonumber
  &=&\frac{2}{\pi}\Gamma(1/3)a^{2/3}\int_0^\infty \d z \sin(z)z^2 \\ \nonumber
  &=&\frac{2}{\pi}\Gamma(1/3)\Gamma(3)a^{2/3} \\ \nonumber
  &\simeq& 8.879 GM n^{2/3},
 \end{eqnarray}
where the third equality follows from the change of variable $u=az^{3/2} F^{-3/2}$ and the definition of the Gamma function, $\Gamma(\alpha)=\int_0^\infty\d x\, x^{\alpha-1} e^{-x}$, whereas the fourth equality results from carrying the integration in complex space $\int \d z\,z^2\sin z={\mathbb Im}[\int \d z\,z^2 \exp(i z)$]. For the experiments shown in Fig.~\ref{fig:evolgen} with $G=M=1$, $N=5\times 10^4$ and $d=10$ we find $\langle F\rangle=46.375$. Note that the typical force has a magnitude $ \langle F\rangle \simeq 2.6 GM /D^2$, where $D$ is the average distance between particles.

A key feature of the Holtsmark distribution is the divergence of the moments $\langle F^\nu\rangle$ for $\nu\ge 2$. This is due to the dominant contribution of the nearest particles to the large-force, power-law tail of the distribution. Using Equation~(\ref{eq:pF_pm_inf}) it is straightforward to show that the variance of $p_0({\bf F})$ diverges in the strong-force limit as
\begin{eqnarray}\label{eq:F2}
  \langle F^2\rangle =\int_0^\infty\d^3 F p_0({\bf F})F^2 \propto  \int_0^\infty \frac{\d F}{ F^{1/2}} \to \infty.
\end{eqnarray}
In practice this means that the maximum force experienced by the test particle can grow up to arbitrarily-large values as the number realizations and time-steps increase in experiment A and B, respectively.

\subsection{Extended substructures}\label{sec:hern}
Self-gravitating substructures with an extended matter distribution generate forces that (i) in general do not diverge at arbitrarily-close distances, and (ii) asymptotically approach the Keplerian limit on scales larger than the size of these systems.  
The simplest modification of Equation~(\ref{eq:fpm}) that accounts for this behaviour is

\begin{eqnarray}\label{eq:fh}
{\bf f}=-\frac{GM}{(r+c)^2} \hat{\bf r}.
\end{eqnarray}
The density profile associated to~(\ref{eq:fh}) can be readily found by solving Poisson's equation, which yields
$$\rho(r)= \frac{M}{2\pi c^3}\frac{1}{(r/c)(1+r/c)^3}.$$
This profile corresponds to a Hernquist (1990) model with a scale length~$c$.\nolinebreak

Let us follow the same steps as in Section~\ref{sec:pm} to derive the spectrum of perturbations induced by a random distribution of Hernquist (1990) spheres.
Inserting Equation~(\ref{eq:fh}) into~(\ref{eq:phi}) and writing ${\bf k}\cdot {\bf f}= k f \cos\theta$ yields

\begin{eqnarray}\label{eq:phih}
  \phi({\bf k})&=& 2\pi n\int_0^{\infty} \d r\, r^2 \int_{-1}^{+1}\d x\big(1-\exp[-i k GMx/(r+c)^2]\big) \\ \nonumber
  &=& 4\pi n \int_0^{\infty} \d r\, r^2 \big(1-\frac{\sin[k GM /(r+c)^2]}{k GM /(r+c)^2}\big)\\ \nonumber
  &=& 4\pi(GM k)^{3/2} n \int_{\psi(k)}^\infty  \d z (z-\psi)^2\big[1- z^2\sin\big(\frac{1}{z^2}\big)\big]\\ \nonumber
  &\equiv& A(k) k^{3/2} ,
\end{eqnarray}
where $\psi(k)=c/\sqrt{GMk}$ is a dimension-less quantity. The last equality follows from the change of variable $u=1/z^2$, with the function $A(k)$ being defined as
\begin{eqnarray}\label{eq:iphih}
A(k)&=&a(I_1+I_2+I_3)~~~~{\rm with} \\ \nonumber 
I_1&=&\frac{1}{2\sqrt{2\pi}}\bigg\{\psi^3\big[-5+2\cos\big(\frac{1}{\psi^2}\big) +
  4\sqrt{2\pi} {\rm FC}\big(\sqrt\frac{2}{\pi}\frac{1}{\psi}\big) \\  \nonumber
 && +   \psi[-4+3b^4\sin\big(\frac{1}{\psi^2}\big) \big] \bigg\}\\ \nonumber
I_2&=&\frac{5\psi^2}{2\sqrt{2\pi}}\bigg\{2\sqrt{2\pi}{\rm FS}\big(\sqrt\frac{2}{\pi}\frac{1}{\psi}\big) \\ \nonumber
&& +\psi\big[-3+2\cos\big(\frac{1}{\psi^2}\big)+\psi^2\sin\big(\frac{1}{\psi^2}\big)\big]\bigg\}\\ \nonumber
I_3&=& -\frac{15\psi}{4\sqrt{2\pi}}\bigg\{\psi^2\big[-2+\cos\big(\frac{1}{\psi^2}\big)+\psi^2\sin\big(\frac{1}{\psi^2}\big)\big]+{\rm Si}\big(\frac{1}{\psi^2}\big)\bigg\}.
\end{eqnarray}
The special functions FS$(x)=\int_0^x\d t \sin(t^2)$, FC$(x)=\int_0^x\d t \cos(t^2)$ correspond to Fresnel sine and cosine integrals, respectively, whereas Si$(x)=\int_0^x \d t\sin(t)/t$ is the sine integral (see e.g. Press et al. 1992 for details). Note that in the limit $c\to 0$ one finds $I_2=I_3=0$, ${\rm FC}(x)\to 1/2$, and $I_1\to 1$, which reduces $A(k)\to a$, thus recovering Equation~(\ref{eq:phipm}).

\begin{figure}
\begin{center}
\includegraphics[width=82mm]{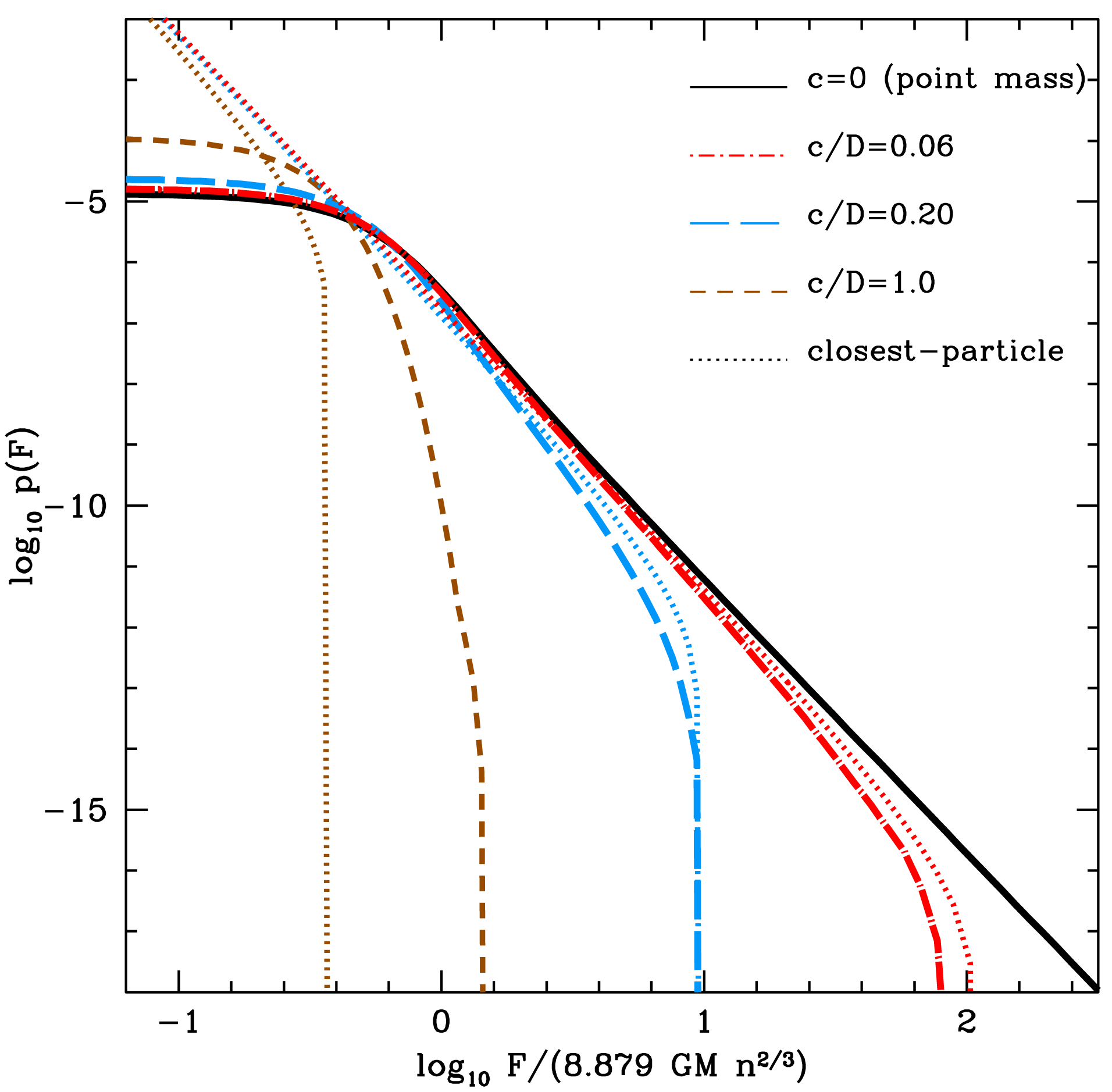}
\end{center}
\caption{Probability density $p({\bf F})$ associated to an homogeneous distribution of equal-mass substructures. The parameter $c$ denotes the scale radius of Hernquist (1990) spheres, whereas $D=(2\pi n)^{-1/3}$ corresponds to the distance at which the probability to find the nearest particle~(\ref{eq:pr_closest}) peaks. The force distribution generated by the closest particle is shown with thin-dotted lines. }
\label{fig:pm_hern}
\end{figure}

Inserting~(\ref{eq:phih}) and~(\ref{eq:iphih}) into~(\ref{eq:pF2}), and following similar steps as in Equation~(\ref{eq:pF_pm}) yields
\begin{eqnarray}\label{eq:pF_hern}
  p({\bf F})=\frac{1}{2\pi^2F}\int_0^\infty \d k\, k \sin(k F) \exp[-A(k) k^{3/2}].
\end{eqnarray}
Fig.~\ref{fig:pm_hern} shows numerical solutions to the integral~(\ref{eq:pF_hern}) for different values of the scale-length $c$. The first noteworthy result is the presence of a truncation of the force distribution at ${\bf F}={\bf F}_0$, such that $p({\bf F})=0$ for ${\bf F}>{\bf F}_0$, which was to be expected given that~(\ref{eq:fh}) is not centrally divergent for $c>0$.

Interestingly, the truncation $F_0$ does {\it not} generally correspond to the maximum force exerted by a single substructure, $f_0=GM/c^2$. To understand this result let us calculate the contribution of the single nearest substructure following the same steps as in Section~\ref{sec:pm}. Recall that for nearby objects $\exp(4\pi n r^3/3)\approx 1$, so that Equation~(\ref{eq:pr_closest}) reduces to $p({\bf r})\d^3 r\approx 4\pi r^2 n \d r$, whereas from Equation~(\ref{eq:fh}) we have $\d r=1/2 (GM)^{1/2}F^{3/2}d F$. It is straightforward to show that the transformation $p({\bf r}) \d^3 r=p_0({\bf F})\d^3 F$, where $|{\bf F}|=G M/(r+c)^2$, yields
\begin{eqnarray}\label{eq:pF_hern_closest}
p_0({\bf F})=\frac{1}{2}(G M)^{3/2}n F^{-9/2}\big(1-\sqrt{F/f_0}\big)^2~~~~ {\rm for}~~~~  F<f_0=\frac{GM}{c^2},
\end{eqnarray}
which recovers Equation~(\ref{eq:p0}) in the limit $f_0\to \infty$.
The function $p_0({\bf F})$  is plotted in Fig.~\ref{fig:pm_hern} with thin-dotted lines.
Note first that for $F\ll f_0$ Equation~(\ref{eq:pF_hern_closest}) greatly overestimates the probability to experience weak forces. The mismatch is caused by the break down of the closest-particle approximation at large distances, where $\exp(4\pi n r^3/3)\ll 1$.

The behaviour of $p({\bf F})$ in the strong-force limit is less clear-cut. Comparison of Equations~(\ref{eq:pF_hern}) and~(\ref{eq:pF_hern_closest}) at $F\sim f_0$ shows that the distribution of forces generated by the nearest substructure either over/under-estimates the collective distribution~(\ref{eq:pF_hern}) depending on whether the scale-length is smaller/larger than the typical separation between substructures, $D=(2\pi n)^{-1/3}$. 
Hence, there are two relevant regimes that define the large-force behaviour of $p({\bf F})$.
\begin{itemize}
\item {\bf Rarefied regime} arises when the separation between substructures is much larger than their individual sizes ($D\gg c$). In this case large forces are dominated by the nearest substructure, such that $F_0\lesssim f_0$.
\item {\bf Saturated regime} arises when substructures overlap with each other ($D\lesssim c$). Here large forces are dominated by the collective contribution of the ensemble rather than by individual objects, hence $F_0\gtrsim f_0$
  \end{itemize}
  The results shown in Fig.~\ref{fig:pm_hern} indicate that the transition between these behaviours occurs on a scale $D\approx 5 c$. In what follows we limit our analysis to rarefied distributions, as an ensemble of self-gravitating, overlapping substructures is not dynamically stable.

\begin{figure}
\begin{center}
\includegraphics[width=84mm]{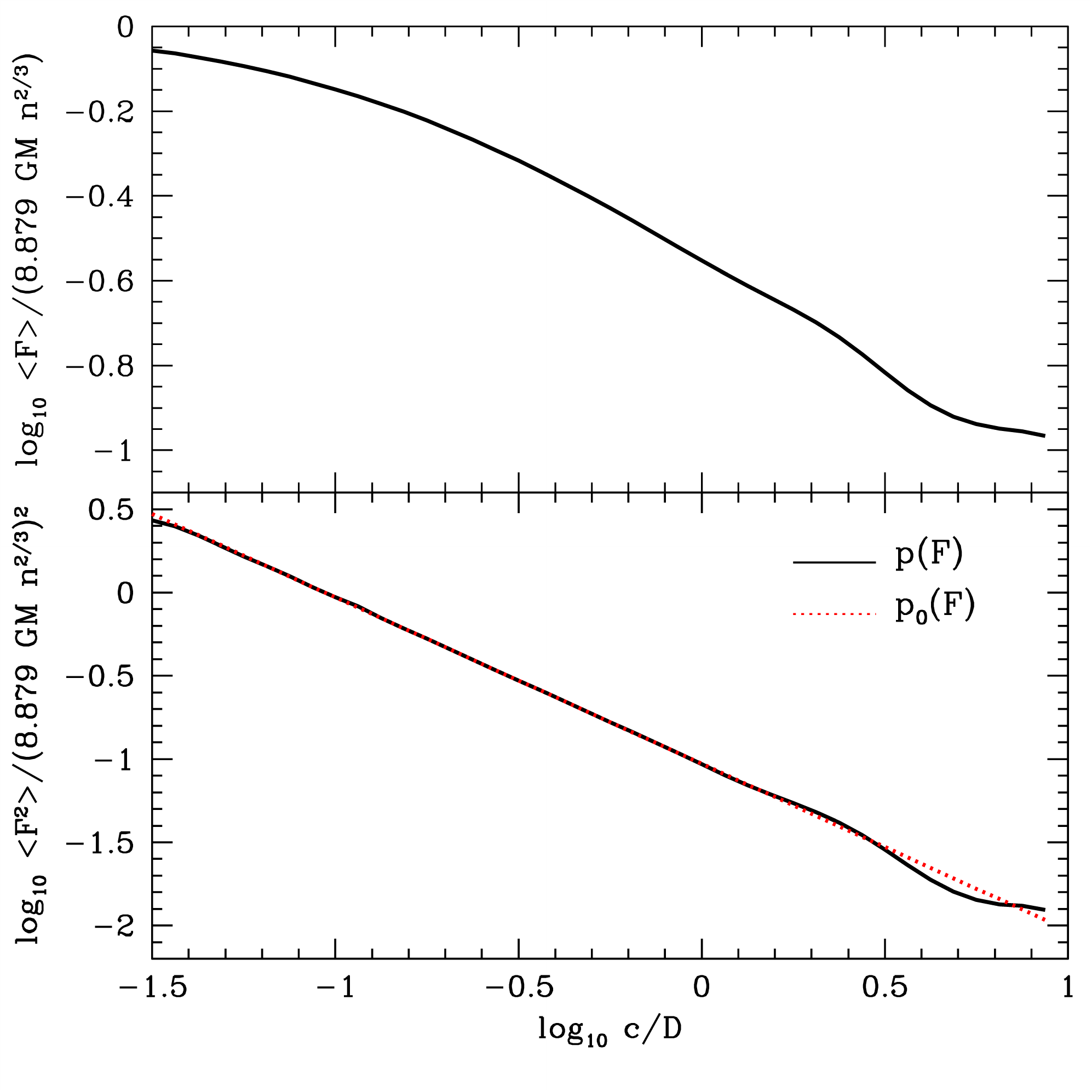}
\end{center}
\caption{First and second moments of the force distribution. {\it Upper panel}: mean specific force $\langle F\rangle$ exerted upon a test particle by a distribution of extended substructures derived from the distribution~(\ref{eq:pF_hern}) as a function of the scale-length $c$, given in units of the average separation between substructures, $D=(2\pi n)^{-1/3}$. Notice the asymptotic convergence towards the value given in Equation~(\ref{eq:F1}) in the limit $c\to 0$. {\it Lower panel}: Variance $\langle F^2\rangle$ calculated from the force distribution~(\ref{eq:pF_hern}) (black-solid lines) and~(\ref{eq:pF_hern_closest}) (red-dotted lines). The good agreement between solid and dotted lines indicates that the force variance is dominated by the closest substructure. }
\label{fig:mom}
\end{figure}

Fig.~\ref{fig:pm_hern} also shows that weak interactions become more likely as the substructure size increases, indicating that the mean specific force and the substructure scale-length are inversely correlated. To inspect this issue in more detail let us calculate $\langle F\rangle$ by inserting~(\ref{eq:pF_hern}) into~(\ref{eq:F1}). In contrast to the average force exerted by a random distribution of point-masses, here $\langle F\rangle$ cannot be expressed analytically owing to the non-trivial functional form of $A(k)$ in~(\ref{eq:iphih}). In the upper panel of Fig.~\ref{fig:mom} we plot numerically-computed values of $\langle F\rangle$ as a function of the substructure scale-length $c$, which is given in units of the typical separation between substructures ($D$). The mean force induced by substructures decreases monotonically as $c/D$ grows. For $c=D$ we find that that $\langle F\rangle$ is a factor $\sim 3$ smaller than an ensemble of point-masses. As expected, for $c\ll D$ the mean force $\langle F\rangle$ converges asymptotically to the value given by Equation~(\ref{eq:F1}).

Crucially, the truncation of the force spectrum~(\ref{eq:pF_hern}) at ${\bf F}={\bf F}_0$ removes the divergence of the force variance discussed in Section~\ref{sec:pm}. This implies that the maximum force experienced by a test particle orbiting in a clumpy medium does {\it not} grow up to arbitrarily-large values insofar as the force generated by individual clumps does not diverge at arbitrarily-close distances.
To illustrate this result let us compute $\langle F^2\rangle$ in the rarefied regime, where nearby particles dominate the large-force tail of Equation~(\ref{eq:pF_hern}), such that $p({\bf F})\approx p_0({\bf F})$. From Equation~(\ref{eq:pF_hern_closest}) 

\begin{eqnarray}\label{eq:F2_hern}
  \langle F^2\rangle &=&\int_0^\infty\d^3 F p_0({\bf F})F^2 \\ \nonumber
  &=&2\pi(G M)^{3/2}n \int_0^{f_0}\frac{\d F}{F^{1/2}}\big(1-\sqrt{F/f_0}\big)^2\\ \nonumber
  &=& \frac{4\pi}{3}(G M)^{3/2}n \sqrt{f_0}\\ \nonumber
  &=&\frac{4\pi}{3}\frac{(G M)^{2}n}{c},
\end{eqnarray}
which shows that the truncation of the large-force spectrum implies a finite variance. As expected, 
the divergence found for point-masses arises in the limit $c\to 0$ ($f_0\to \infty$).
The close agreement between the variance derived from Equations~(\ref{eq:pF_hern}) and~(\ref{eq:pF_hern_closest}) visible in the lower panel of Fig.~\ref{fig:mom} suggests that, as in the case of point-mass particles, nearby substructures dominate the large-force, power-law tail of the force distribution.

\begin{figure*}
\begin{center}
\includegraphics[width=174mm]{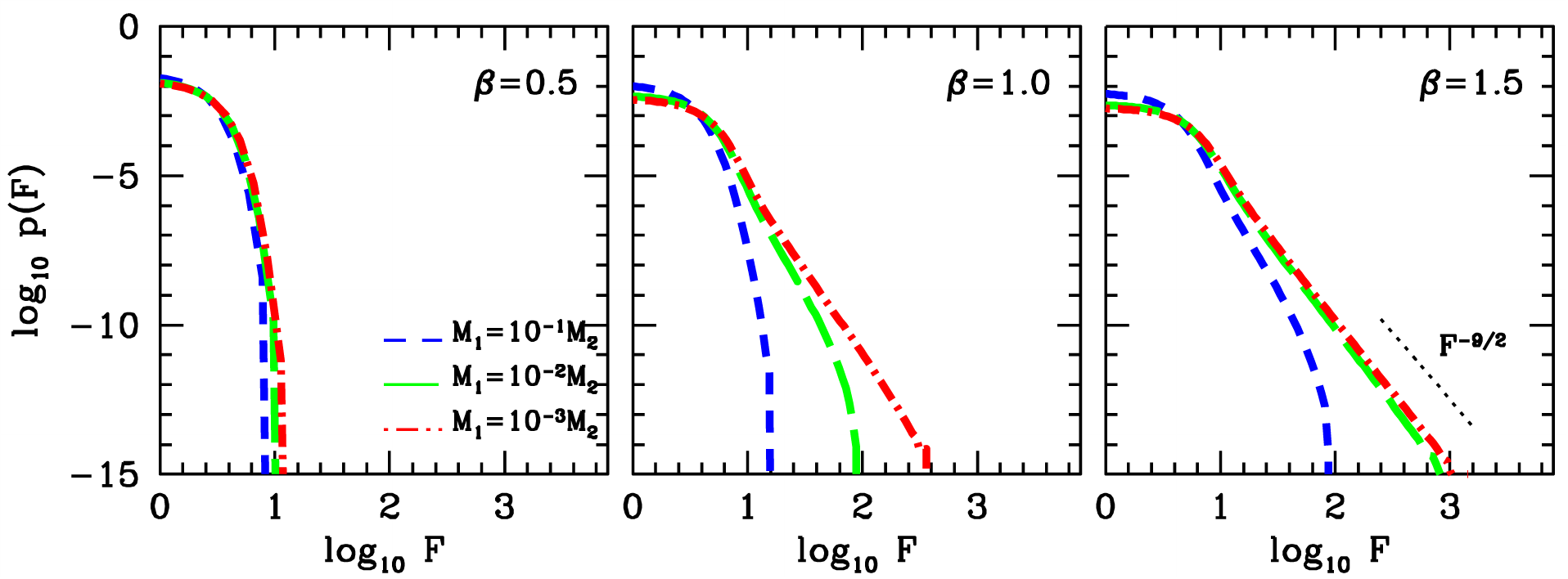}
\end{center}
\caption{Force distribution arising from a population of extended substructures with a power-law mass function~(\ref{eq:pl}), index $\alpha=-2$ and normalization $B_0=1$. The upper-mass limit is fixed to $G=M_2=M_0=1$, and the lower-mass limit $M_1$ is given as a free parameter. Each of the three panels corresponds to a different index of the power-law relation between the scale-radius and the mass of substructures, $c(M)=c_0(M/M_0)^\beta$, with $c_0=1$. In particular, left, middle and right panels adopt power-law indices $\beta=0.5, 1.0$ and 1.5, respectively. Note that for $\beta\ge 1$ the large-force tail of the distribution approaches the point-mass tail  $p({\bf F})\sim F^{-9/2}$ in the limit $M_1\to 0$. In this regime force fluctuations can reach arbitrarily-large values (see \S\ref{sec:pm}).}
\label{fig:pf_pl}
\end{figure*}

\subsection{Substructures with a mass function}\label{sec:massfun}
The above results can be straightforwardly extended to ensembles of substructures with a mass distribution by writing the number density as
$$n\to \int \frac{\d n}{\d M}\d M.$$
Power-law mass functions are of particular interest 
\begin{eqnarray}\label{eq:pl}
\frac{\d n}{\d M}=B_0\bigg(\frac{M}{M_0}\bigg)^\alpha,
\end{eqnarray}
where $M_0$ is our mass unit, $B_0$ a normalization parameter and $\alpha$ the power-law index. In general, the mass function is defined within a range $(M_1,M_2)$, with $M_2> M_1$. To normalize the distribution we impose the condition $\int_V \d^3r \int_{M_1}^{M_2}\d M (\d n/\d M)= N$, where $N$ is the total number of substructures within a volume $V=4\pi d^3/3$. Hence,
\begin{eqnarray}\label{eq:B0}
  B_0=n_0M_0^\alpha\times
  \begin{cases}
    \frac{1+\alpha}{M_2^{1+\alpha}-M_1^{1+\alpha}} & ,\alpha\ne - 1 \\
    \frac{1}{\ln(M_2/M_1)} & ,\alpha= - 1,
    \end{cases}
\end{eqnarray}
where $n_0=N/V$. Below we examine the impact of the mass function~(\ref{eq:pl}) on the substructure models outlined in Sections~\ref{sec:pm} and~\ref{sec:hern} separately.

\subsubsection{Point-mass particles}\label{sec:pm_mf}
Let us first calculate the function $\phi({\bf k})$ by taking into account that particles in the mass range $M,M+\d M$ contribute to the number density by an amount $\d n$. Hence, from Equation~(\ref{eq:phipm}) the total contribution of particles with masses between $(M_1,M_2)$ is
\begin{eqnarray}\label{eq:phipm_pl}
  \phi({\bf k})&=&\frac{4}{15}(2\pi G)^{3/2}B_0 k^{3/2} \int_{M_1}^{M_2}\d M\, M^{3/2}\bigg(\frac{M}{M_0}\bigg)^\alpha \\ \nonumber
  &=&k^{3/2}\frac{4}{15}(2\pi G)^{3/2}\frac{B_0}{M_0^\alpha} \times
\begin{cases}
\frac{M_2^{\alpha+5/2}-M_1^{5/2+\alpha}}{\alpha+5/2}  & ,\alpha\ne - 5/2 \\ 
\ln\big( {M_2/M_1}\big)  & , \alpha=-5/2, 
\end{cases}
\\ \nonumber
&\equiv& k^{3/2} a'.
\end{eqnarray}
Thus, varying the parameters of the mass function simply modifies the numerical value of $a'$.

Inserting $a'$ and~(\ref{eq:B0}) into~(\ref{eq:F1}) yields
\begin{eqnarray}\label{eq:F1_pl}
    \langle F\rangle &=&\frac{2}{\pi}\Gamma(1/3)\Gamma(3)a'^{2/3} \\ \nonumber
  &\simeq& 8.879 G n_0^{2/3}\bigg(\frac{\alpha+1}{\alpha+5/2} \frac{M_2^{\alpha+5/2}-M_1^{\alpha+5/2}}{M_2^{\alpha+1}-M_1^{\alpha+1}}\bigg)^{2/3},
\end{eqnarray}  
where the indices $\alpha= -1$ and $\alpha= -5/2$ have been excluded for clarity. Hence, for $M_2\gg M_1$ we find that the average force scales as
\begin{itemize}
\item $\langle F\rangle \sim G M_2 n_0^{2/3}$ for $\alpha>-1$ (shallow mass profile).
\item $\langle F\rangle \sim G M_2^{(2\alpha+5)/3}M_1^{2(\alpha+1)/3}n_0^{2/3}$ for $-5/2<\alpha<-1$.
  \item $\langle F\rangle \sim G M_1n_0^{2/3}$ for $\alpha<-5/2$ (steep mass profile).  
  \end{itemize}

Similarly, replacing $a\to a'$ in Equation~(\ref{eq:pF_pm_inf}) and inserting the resulting distribution into~(\ref{eq:F2}) yields a divergent $\langle F^2\rangle $ independently of the value of $\alpha$. As discussed in Section~\ref{sec:pm}, the divergence of the second moment of $p({\bf F})$ can be traced back to the singular behaviour of the Keplerian force~(\ref{eq:fpm}) at $r=0$.

\subsubsection{Extended substructures}\label{sec:hern_mf}
Self-gravitating substructures in dynamical equilibrium have sizes that correlate with their masses. It is useful to quantify this dependency by introducing a size function $c(M)$, which determines the variation of the scale-radius as a function of $M$. The power-law mass function seen in Section~\ref{sec:pm_mf} can be readily modified in order to account for the mass-size relation by writing
$$n\to \int\int \frac{\d^2 n}{\d M\d c}\d M\d c$$
 and
\begin{eqnarray}\label{eq:nmc}
\frac{\d^2 n}{\d M\d c}=B_0\bigg(\frac{M}{M_0}\bigg)^\alpha\delta[c-c(M)].
\end{eqnarray}
where $\delta$ denotes Dirac's delta function. Equation~(\ref{eq:phih}) then becomes
\begin{eqnarray}\label{eq:phih_pl}
  \phi({\bf k})
 & =&4\pi G^{3/2} B_0k^{3/2} \int_{M_1}^{M_2}\d M M^{3/2}\bigg(\frac{M}{M_0}\bigg)^\alpha\\ \nonumber
 && \times \int_{\psi_M(k)}^\infty  \d z (z-\psi_M)^2\big[1- z^2\sin\big(\frac{1}{z^2}\big)\big] \\ \nonumber
  &\equiv& A'(k) k^{3/2},
\end{eqnarray}
where $\psi_M(k)=c(M)/\sqrt{G M k}$, and $A'(k)$ is defined as
\begin{eqnarray}\label{eq:iphih_pl}
A'(k)=\frac{4}{15}(2\pi G)^{3/2}\frac{B_0}{M_0^\alpha}\int_{M_1}^{M_2} \d M M^{\alpha+3/2}\sum_{j=1}^3I_j(k,M).
\end{eqnarray}
with the $I_j$ functions given by Equation~(\ref{eq:iphih}). 

The relation between mass \& scale-radius of extended substructures, $c(M)$, plays a key role in defining the large-force behaviour of $p({\bf F})$. In particular, it determines whether the strongest forces arise from the most massive objects or, in contrast, from the smallest, most abundant substructures. 
To study this issue in some detail let us adopt a power-law relation
\begin{eqnarray}\label{eq:cm}
  c(M)=c_0\bigg(\frac{M}{M_0}\bigg)^\beta.
\end{eqnarray}
Given that in the rarefied regime the variance of the force distribution is dominated by the contribution of the nearest substructure, we can use Equations~(\ref{eq:F2_hern}),~(\ref{eq:pl}) and~(\ref{eq:cm}) to derive an analytic expression for the force variance, which yields
\begin{eqnarray}\label{eq:F2_pl}
  \langle F^2\rangle&=& \frac{4\pi}{3}\int_{M_{1}}^{M_2}\d M\frac{\d n}{\d M}\frac{G^2 M^{2}}{c(M)}\\ \nonumber
  &=& \frac{4\pi}{3}M_0^{\beta-\alpha}\frac{G^2 B_0}{c_0}\times
  \begin{cases}
    \frac{M_2^{3+\alpha-\beta}-M_1^{3+\alpha-\beta}}{3+\alpha-\beta} &,~ 3+\alpha-\beta \ne 0 \\
    {\ln(M_2/M_1)} & ,~3+\alpha-\beta = 0,
    \end{cases}
\end{eqnarray}
Note that for ensembles with a broad mass range, $M_2\gg M_1$, Equation~(\ref{eq:F2_pl}) exhibits two well-defined regimes
\begin{itemize}
\item {\bf Macrostructure-dominated force distribution}. This regime corresponds to a size function with index $\beta < 3+\alpha$, such that $M_2^{3+\alpha-\beta}-M_1^{3+\alpha-\beta}\simeq M_2^{3+\alpha-\beta}$, yielding a variance $\langle F^2\rangle\sim M_2^{3+\alpha-\beta}$ dominated by the largest objects.
\item {\bf Microstructure-dominated force distribution}. In this case the least-massive substructures govern the large-force behaviour of $p({\bf F})$. This regime arises when the size function of substructures is sufficiently steep, $\beta \ge 3+\alpha$, which leads to a force distribution whose variance scales as $\langle F^2\rangle\sim M_1^{3+\alpha-\beta}$, thus diverging in the limit $M_1\to 0$.
  This is an important result, as it shows that the probability of experiencing large force fluctuations strongly depends on the shape of the size function $c(M)$ on mass scales $M_1\lll M_2$.
 \end{itemize}

\subsubsection{Example: cosmological index $\alpha=-2$}\label{sec:cosmo_pf}
It is useful to illustrate the above results with an idealized representation of the cosmological $N$-body models studied in Section~\ref{sec:micro}, which follow a power-law mass function~(\ref{eq:pl}) with an index $\alpha\simeq -2$ (e.g. Springel et al. 2008), a fixed normalization parameter $B_0$, and a lower mass limit $M_1$ that contains information on the quantum attributes of the dark matter particle candidates.

Notice first that a mass function with index $\alpha=-2$ within the interval $(M_1,M_2)$ contains a divergent number of substructures in the limit $M_1\to 0$. Indeed, 
integration of~(\ref{eq:pl}) yields $N\simeq V B_0 M_0^2 M_1^{-1}$, which grows to arbitrarily-large values as the minimum mass of substructures decreases.

The average force exerted by an ensemble of point-mass particles with $\alpha=-2$ is completely dominated by the most massive objects. This 
can be straightforwardly shown from Equation~(\ref{eq:F1_pl})
\begin{eqnarray}\label{eq:F1_pl_2}
  \langle F\rangle
  &\simeq& 8.879 G n_0^{2/3}\bigg(2\frac{M_2^{1/2}-M_1^{1/2}}{ M_1^{-1}-M_2^{-1}}\bigg)^{2/3}\\ \nonumber
  &\simeq& 14.095 G B_0^{2/3}M_0^{4/3}M_2^{1/3},
\end{eqnarray}
where we have substituted $n_0=B_0 M_0^2 M_1^{-1}$. Note that the mean force scales as $\langle F\rangle \sim M_2^{1/3}$ independently of the lower limit of the mass function, $M_1$. This is noteworthy given that, in principle, the number of light particles can become arbitrarily large.

As discussed in \S\ref{sec:hern_mf}, the variance of the force distribution depends on the shape of the size function $c(M)$. Substituting $\alpha=-2$ in Equation~(\ref{eq:F2_pl}) yields
\begin{eqnarray}\label{eq:F2_pl_2}
  \langle F^2\rangle =\frac{4\pi}{3}M_0^{2+\beta}\frac{G^2 B_0}{c_0} \frac{M_2^{1-\beta}-M_1^{1-\beta}}{1-\beta}.
\end{eqnarray}
Hence, for $\beta< 1$ the numerator in Equation~(\ref{eq:F2_pl_2}) can be approximated as $M_2^{1-\beta}-M_1^{1-\beta}\approx M_2^{1-\beta}$, which shows that the variance is dominated by the most massive objects, whereas for $\beta> 1$ the variance $\langle F^2\rangle \sim M_1^{1-\beta}$ diverges in the limit $M_1\to 0$, indicating that the smallest substructures induce the largest fluctuations.

To gain further insight onto this result we show in 
Fig.~\ref{fig:pf_pl} the force distribution $p({\bf F})$ arising from a population of extended substructures with a power-law mass function $\alpha=-2$. To calculate $p({\bf F})$ we insert Equations~(\ref{eq:iphih_pl}) and~(\ref{eq:cm}) in~(\ref{eq:pF_hern}) and integrate over $\d k$. For simplicity we adopt $M_2=M_0=B_0=c_0=1$. Also, for ease of comparison with previous Sections the volume is set to $V=4\pi d^3/3$ with $d=10$, which leaves $M_1$ as the only free parameter.
Each panel corresponds to a different value of the index $\beta$ in the size distribution~(\ref{eq:cm}). Adopting $\beta=0.5$ in the left panels reveals a scant sensitivity of $p({\bf F})$ to the lower-mass limit, $M_1$. Indeed, this is a macrostructure-dominated distribution, where the mean and the variance of $p({\bf F})$ are largely determined by the most massive substructures in the ensemble. In contrast, for power-law indices $\beta= 1.0$ and 1.5 (middle and right panels respectively) the large-force distributions are microstructure-dominated, exhibiting a strong dependence on the lower mass limit, $M_1$. Notice in particular that for steep size functions ($\beta\ge 1$) the large-force tail of the force distribution approaches the point-mass behaviour $p({\bf F})\sim F^{-9/2}$ in the limit $M_1\to 0$, which according to Equation~(\ref{eq:F2}) leads to a divergent variance. 

Interestingly, CDM subhaloes found in collision-less $N$-body simulations of structure formation exhibit power-law size functions with $\beta\lesssim 1$. For example, using the Via Lactea II models (Diemand et al. 2008) Erkal et al. (2016) find that $\beta\simeq 0.5$ provide the best fit to the mass-size relation at redshift $z=0$.
Note, however, that cosmological simulations of Milky Way-size haloes typically resolve substructures with masses $10^{6} \lesssim M/M_\odot\lesssim 10^{10}$, with the low-mass limit only partially imposed by the finite mass resolution of the $N$-body runs (see van den Bosch 2017).
Thus, (boldly) extrapolating these results $\sim 12$ orders in magnitude in mass, down to the mass scale associated with the streaming length of CDM particles ($M_1\sim 10^{-6}M_\odot$, e.g. S{\'a}nchez-Conde \& Prada 2014; Molin\'e et al. 2017 and references therein) suggests a negligible contribution of the smallest dark matter mini-haloes to the force distribution $p({\bf F})$. We will return to this issue in \S\ref{sec:dis}.

\section{Stochastic tidal forces}\label{sec:tides}
The equations of motion that govern the phase-trajectory of a tracer particle ${\bf R}'(t)$ within a self-gravitating system that moves through a clumpy medium can be written as 
\begin{eqnarray}\label{eq:eqmots}
\frac{{\d^2 \bf R}'}{\d t^2}=-\nabla\Phi_s({\bf R}') + T_g\cdot {\bf R}' + \sum_{i=1}^{N} t_i\cdot {\bf R}',
\end{eqnarray}
where $T_g$ and ${t}_i$ are 3$\times$3 tidal tensors evaluated at the centre of the self-gravitating potential $\Phi_s$. The smooth component has a form
\begin{eqnarray}\label{eq:tt_s}
  {T}_{g}^{jk}\equiv -\frac{\partial^2 \Phi_g}{\partial x_j\partial x_k}, 
\end{eqnarray}
while the stochastic tidal tensor  
\begin{eqnarray}\label{eq:tt}
 {T}^{jk}\equiv  \sum_{i=1}^{N}{t}_i= \sum_{i=1}^{N} \frac{\partial f^k_{i}}{\partial x^j}=\frac{\partial }{\partial x^j}\sum_{i=1}^{N} f^k_{i}=\frac{\partial F^k}{\partial x^j},
\end{eqnarray}
arises from the gradient in the combined tidal force induced by a set of $N-${substructures} distributed across the external galaxy.

The tidal force induced by a single substructure can be written as
\begin{eqnarray}\label{eq:ftid}
  {\bf f}_{t}\equiv t\cdot {\bf R}'.
\end{eqnarray}
To simplify the analysis it is useful to diagonalize the tensor ${t}$ by rotating the coordinates to a new frame $\d{\bf R}/\d t=\d {\bf R}'/\d t-{\bb \Omega}\times {\bf R}$, which co-rotates with the angular velocity of the substructure, ${\bb \Omega}$. In the non-inertial frame the 
effective tensor $t_e=t+ {\bf \nabla f}_c$ has a diagonal form, where ${\bf f}_c={\bb \Omega}\times ({\bb \Omega}\times {\bf R})$ is the centrifugal force component (see Renaud et al. 2011 for details). Unfortunately, the resulting field is not spherically isotropic, which greatly complicates the mathematical analysis of random forces outlined in Section~\ref{sec:fluctu}. Yet, at leading order one can isotropize the tidal field by writing Equation~(\ref{eq:ftid}) as 
\begin{eqnarray}\label{eq:ftid2}
  {\bf f}_t\approx R\lambda  \hat{\bf u}\equiv R{\bb \lambda},
  \end{eqnarray}
where 
$\lambda ={\rm Trace}( {t}_e)$ is the sum of eigenvalues of the effective tidal tensor $t_e$, and $\hat {\bf u}$ is a unit vector irrelevant for our analysis (see \S\ref{sec:plam_pm}). 
Equation~(\ref{eq:ftid2}) also neglects the Euler and Coriolis terms appearing in the non-inertial rest frame, implicitly assuming that during the duration of the encounter the angular frequency ${\bb \Omega}$ remains constant and in a parallel direction to the galactocentric velocity $\d{\bf R}/\d t$, respectively. This approximation is less inaccurate in an impulsive regime, where the time-scale of the tidal fluctuations is much shorter than the orbital period of the tracer particle in the potential $\Phi_s$.

 With these approximations in place it is relatively straightforward to extend the analysis of Section~\ref{sec:fluctu} to systems experiencing stochastic tidal perturbations. In analogy with Equation~(\ref{eq:F}) let us first define the vector 
\begin{eqnarray}\label{eq:Lambda}
  {\bb \Lambda}\equiv \sum_{i=1}^N {\bb \lambda}_i.
\end{eqnarray}
which arises from the combined tidal field of the external substructures. From~(\ref{eq:ftid2}) and~(\ref{eq:Lambda}) the combined tidal force can be written as
\begin{eqnarray}\label{eq:Flambda}
    {\bf F}_t=\sum_{i=1}^N {\bb \lambda}_iR=\bb\Lambda R.
\end{eqnarray}
A tracer particle embedded in a sea of substructures isotropically distributed in space experiences a force ${\bf F}_t$ that fluctuates in random directions according to a probability function $p({\bf F}_t)$, which is fully specified by the tidal distribution $p({\bb \Lambda})$. Below we derive this function for isotropic ensembles of point-mass particles and Hernquist (1990) spheres.

\subsection{Point-mass particles}\label{sec:plam_pm}
Consider a vector
\begin{eqnarray}\label{eq:lambdapm}
\bb \lambda=\frac{2GM}{r^3}\hat{\bf u},
\end{eqnarray}
where $|{\bb \lambda}|$ corresponds to the trace of the tidal tensor induced by an external Keplerian force~(\ref{eq:fpm}) (see Equation~(18) of Renaud et al. 2011).

Inserting Equation~(\ref{eq:lambdapm}) into~(\ref{eq:phi}) and integrating over $\d^3r=2\pi r^2\d r\d\cos(\theta)$, with ${\bf k}\cdot{\bb \lambda}=k \lambda \cos(\theta)$, yields
\begin{eqnarray}\label{eq:philampm}
  \phi({\bf k})&=& 2\pi n\int_0^{\infty} \d r\, r^2 \int_{-1}^{+1}\d x\big(1-\exp[-i k 2GMx/r^3]\big) \\ \nonumber
  &=& 2\pi GMn k \int_0^{\infty} \d z \,z^2 \big[2-z^3\sin\big(\frac{2}{z^3}\big)\big]\\ \nonumber
  &=&\frac{2\pi^2}{3}GMn k\equiv q k,
\end{eqnarray}
with $q\equiv \frac{2\pi^2}{3}GMn\simeq GM/D^3$.

The distribution $p({\bb \Lambda})$ can be expressed in an analytical form. Inserting~(\ref{eq:philampm}) into~(\ref{eq:pF2}) and integrating over $\d^3 k$ yields
\begin{eqnarray}\label{eq:plam_pm}
  p({\bb \Lambda})&=&\frac{1}{(2\pi)^2}\int_0^\infty \d k\, k^2\exp(-q k)\int_{-1}^{+1}\d x\exp[-i k \Lambda x]\\ \nonumber 
  &=&\frac{1}{2\pi^2}\int_0^\infty \d k\, k^2 \exp(-q k)\frac{\sin(k \Lambda)}{k\Lambda} \\ \nonumber
  &=&\frac{1}{\pi^2q ^3}\frac{1}{(1+\xi^2)^2},
\end{eqnarray}
where $\xi\equiv \Lambda/q$ is a dimension-less quantity.

The distribution~(\ref{eq:plam_pm}) exhibits two well-defined asymptotic behaviours depending on the magnitude of $\xi$. In the {\it weak}-tides limit ($\xi\ll 1$) Equation~(\ref{eq:plam_pm}) becomes flat
\begin{eqnarray}\label{eq:plam_weak}
 \lim_{\Lambda\to 0} p({\bb \Lambda})= \frac{27}{8\pi^8}(G M n)^{-3},
\end{eqnarray}
whereas in the {\it strong}-tides limit ($\xi\gg 1$) it follows a power-law tail
\begin{eqnarray}\label{eq:plam_strong}
 \lim_{\Lambda\to \infty}   p({\bb \Lambda})=  \frac{2}{3}\frac{GM n}{\Lambda^4}.
\end{eqnarray}

As in Section~\ref{sec:pm}, the large-force behaviour is entirely dominated by the nearest particle. Indeed, from Equation~(\ref{eq:pr_closest}) the probability to find the closest particle at a distance ${\bf r}$ is $p({\bf r})\simeq 4\pi n r^2\d r$, while~(\ref{eq:lambdapm}) leads to $r=(2 G M/\Lambda)^{1/3}$. Hence, from~(\ref{eq:p0}) one has that
\begin{eqnarray}\label{eq:p0lambda}
  p_0({\bb \Lambda})=n\frac{r^2\d r}{\Lambda^2\d \Lambda}=\frac{2}{3}\frac{G M n}{ \Lambda^4},
   \end{eqnarray}
  which matches the strong-tide limit~(\ref{eq:plam_strong}) exactly.

It is straightforward to show that both the mean $\langle\Lambda\rangle$ and the variance $\langle\Lambda^2\rangle$ of the distribution $p({\bb \Lambda})$ diverge in the limit $\Lambda\to \infty$, that is when the distance to the closest particle becomes arbitrarily small. Thus, the divergence of the tidal moments is caused by the singular behaviour of~(\ref{eq:lambdapm}) at $r=0$.

\begin{figure}
\begin{center}
\includegraphics[width=82mm]{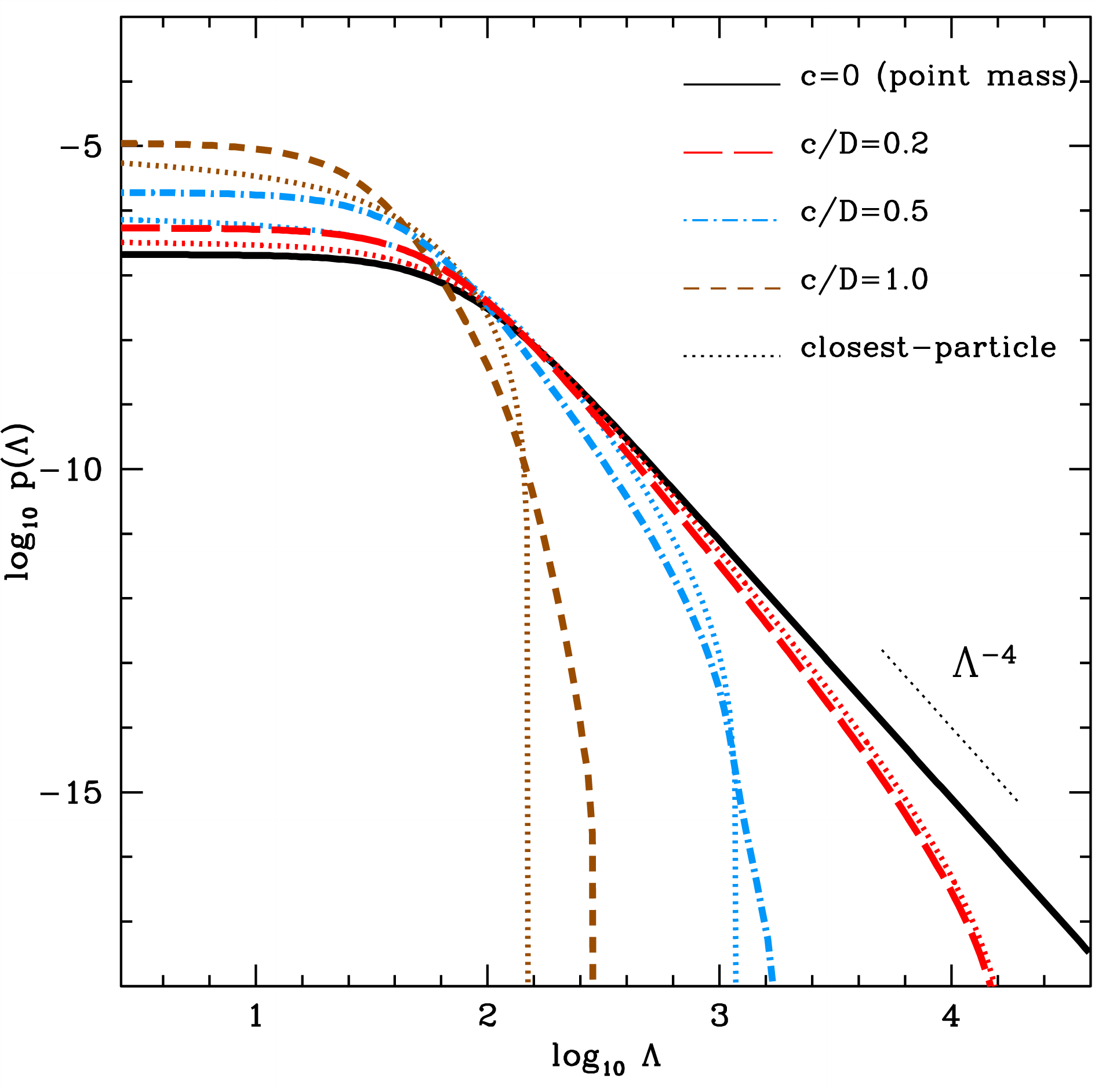}
\end{center}
\caption{Tidal force distribution $p({\bb \Lambda})$ associated to an homogeneous distribution of equal-mass substructures with different scale radii, $c$, given in units of $D=(2\pi n)^{-1/3}$, which corresponds to the distance at which the probability to find the nearest particle~(\ref{eq:pr_closest}) is highest. For ease of comparison with Fig.~\ref{fig:evolgen} we choose $N=5\times 10^4$ particles with a mass $M=G=1$ distributed over a volume $V=4\pi d^3/3$ with $d=10$. The analytical distribution~(\ref{eq:plam_hern_1}) is shown with thin-dotted lines. }
\label{fig:pm_hern_lam}
\end{figure}

\subsection{Extended substructures}\label{sec:plam_hern}
The trace vector associated with a Hernquist (1990) sphere can be straightforwardly derived from Equation~(17) of Renaud et al. (2011), which yields
\begin{eqnarray}\label{eq:lambdah}
\bb \lambda=\frac{2GM}{(r+c)^3}\hat{\bf u}.
\end{eqnarray}
As expected, Equation~(\ref{eq:lambdah}) approaches asymptotically the point-mass behaviour~(\ref{eq:lambdapm}) at distances much larger than the size of the substructure, $r\gg c$.

We can now follow the same procedure as in Section~\ref{sec:hern} to derive the distribution $p({\bb \Lambda})$ of tidal forces generated by random ensembles of Hernquist (1990) spheres.
Inserting Equation~(\ref{eq:lambdah}) into~(\ref{eq:phi}) and writing ${\bf k}\cdot {\bb \lambda}= k \lambda \cos\theta$ yields

\begin{eqnarray}\label{eq:phih_l}
  \phi({\bf k})&=& 2\pi n\int_0^{\infty} \d r\, r^2 \int_{-1}^{+1}\d x\big(1-\exp[-i k 2GMx/(r+c)^3]\big) \\ \nonumber
  &=& 2\pi n \int_0^{\infty} \d r\, r^2 \big(2-\frac{\sin[k 2GM /(r+c)^3]}{k GM /(r+c)^3}\big)\\ \nonumber
  &=& 2\pi(GM k) n \int_{\varphi(k)}^\infty  \d z (z-\varphi)^2\big[2- z^3\sin\big(\frac{2}{z^3}\big)\big]\\ \nonumber
  &\equiv& Q(k) k ,
\end{eqnarray}
where $\varphi(k)=c/(GMk)^{1/3}$ is a dimension-less quantity, and the function $Q(k)$ is defined as
\begin{eqnarray}\label{eq:iphih_l}
Q(k)&=&q(J_1+J_2+J_3)~~~~{\rm with} \\ \nonumber 
J_1&=&\frac{1}{2\pi}\bigg\{\varphi^3\big[-4+2\cos\big(\frac{2}{\varphi^3}\big)+\varphi^3\sin\big(\frac{2}{\varphi^3}\big)\big] +4 {\rm Si}\big(\frac{2}{\varphi^3}\big) \bigg\}\\ \nonumber
J_2&=&\frac{3\varphi^2}{4\pi}\bigg[-8\varphi+6\varphi\cos\big(\frac{2}{\varphi^3}\big) -\frac{12}{\varphi^2}{\rm Si}\big(\frac{1}{3},\frac{2}{\varphi^3}\big)\\ \nonumber
&&+3^{3/2} 2^{1/3}\Gamma\big(2/3\big) +\varphi^4\sin\big(\frac{2}{\varphi^3}\big) \bigg]\\ \nonumber
J_3&=& -\frac{3}{5\pi}\bigg[-10\varphi^3+6\varphi^3\cos\big(\frac{2}{\varphi^3}\big)-12 {\rm Si}\big(\frac{2}{3},\frac{2}{\varphi^3}\big)\\ \nonumber
&&+3\cdot 2^{2/3}\varphi\Gamma\big(1/3\big) +2\varphi^6 \sin\big(\frac{2}{\varphi^3}\big)\bigg],
\end{eqnarray}
where  ${\rm Si}(n,x)=\int_1^\infty\d t \sin(x t)/t^{n}$ is the generalized sine integral (Press et al. 1992). After some algebra one can readily show that for $\varphi\lesssim 1$ Equation~(\ref{eq:iphih_l}) scales as
\begin{eqnarray}\label{eq:iphih2_l}
Q(k)=q\bigg[1-\frac{2^{2/3}9\Gamma(1/3)}{5\pi}\varphi+\frac{3^{5/2}\Gamma(2/3)}{2^{5/3}\pi}\varphi^2-\frac{2}{\pi}\varphi^3 +\mathcal{O}(\varphi^4)\bigg]
\end{eqnarray}
such that $\lim_{c\to 0} Q(k)=q$, thus recovering the point-mass limit discussed in \S\ref{sec:plam_pm}.

We can now follow similar steps as in Equation~(\ref{eq:pF_pm}) to derive the tidal force distribution
\begin{eqnarray}\label{eq:pL_hern}
  p({\bb \Lambda })=\frac{1}{2\pi^2\Lambda}\int_0^\infty \d k\, k \sin(k \Lambda) \exp[-Q(k) k],
\end{eqnarray}
which is isotropically oriented in space, i.e. $p({\bb \Lambda })=p({ \Lambda })$.
Solving~(\ref{eq:pL_hern}) becomes a numerical challenge for scale radii $c\ll D$, i.e. when the separation between substructures is much larger than their individual sizes. As shown in Section~\ref{sec:hern}, this corresponds to a rarefied regime of perturbations, where the large-force behaviour of $p({\bb \Lambda })$ is entirely dominated by the nearest substructure. In this regime one can use Equation~(\ref{eq:pr_closest}) to derive the large-force tail of the distribution by writing the probability to find the closest particle at a distance as ${\bf r}$ as $p({\bf r})\simeq 4\pi n r^2\d r$, whereas from~(\ref{eq:lambdah}) one has that $r+c=(2 G M/\Lambda)^{1/3}$, which therefore yields
\begin{eqnarray}\label{eq:p0lambdah}
 p_0({\bb \Lambda})=n\frac{r^2\d r}{\Lambda^2\d \Lambda}=\frac{2}{3}\frac{G M n}{ \Lambda^4}\big[1-\big(\Lambda/\lambda_0\big)^{1/3}\big]^2 ~~~~~~{\rm for}~~\Lambda< \lambda_0.
   \end{eqnarray}
As discussed in Section~\ref{sec:hern}, the fact that Equation~(\ref{eq:lambdah}) is not centrally divergent introduces a truncation in the force distribution~(\ref{eq:p0lambdah}) at $\Lambda\approx \lambda_0$, which for $c\ll D$ roughly corresponds to the maximum force exerted by an individual substructure, $\lambda_0\approx 2GM/c^3$.

Comparison of~(\ref{eq:p0lambdah}) and~(\ref{eq:p0lambda}) shows that the truncation has an algebraic form $g(x)=(1-x^{1/3})^2$ for $x< 1$, and $g(x)=0$ otherwise. Hence, in the rarefied regime of perturbations we can derive an approximate analytical expression for $p({\bb \Lambda})$ by as $p_1({\bb \Lambda })=p_{c=0}({\bb \Lambda })g(\Lambda/\lambda_0)$, where $p_{c=0}({\bb \Lambda })$ is the point-mass distribution~(\ref{eq:plam_pm}), such that
\begin{eqnarray}\label{eq:plam_hern_1}
  p_1({\bb \Lambda})= \frac{C}{\pi^2q^3}\frac{1}{(1+\xi^2)^2}\big[1-\big(q/\lambda_0\big)^{1/3}\xi^{1/3}\big]^2~~~~~~{\rm for}~~\Lambda< \lambda_0,
\end{eqnarray}
where $\xi= \Lambda/q$ is the dimension-less quantity defined in \S\ref{sec:plam_pm}, and $C$ is a normalization parameter that guarantees $\int_0^{\lambda_0}\d^3\Lambda\,p_1({\bb \Lambda})=1$. Comparison between~(\ref{eq:plam_hern_1}) and (\ref{eq:plam_pm}) indicates that $C\to 1$ in the limit $\lambda_0\to \infty$ ($c\to 0$). 
To inspect the accuracy of this approximation we plot in Fig.~\ref{fig:pm_hern_lam} the distributions~(\ref{eq:pL_hern}) and~(\ref{eq:plam_hern_1}) with thick and thin-dotted lines, respectively, for various values of the scale radius of the Hernquist (1990) spheres ($c$). The solid line denotes the point-mass function~(\ref{eq:plam_pm}), which falls off as $p_{c=0}({\bb \Lambda })\sim \Lambda^{-4}$ at $\Lambda\to \infty$. As expected, the agreement between $p({\bb \Lambda})$ and $p_1({\bb \Lambda})$ improves as the value of $c$ decreases with respect to the average separation between substructures.

\begin{figure*}
\begin{center}
\includegraphics[width=174mm]{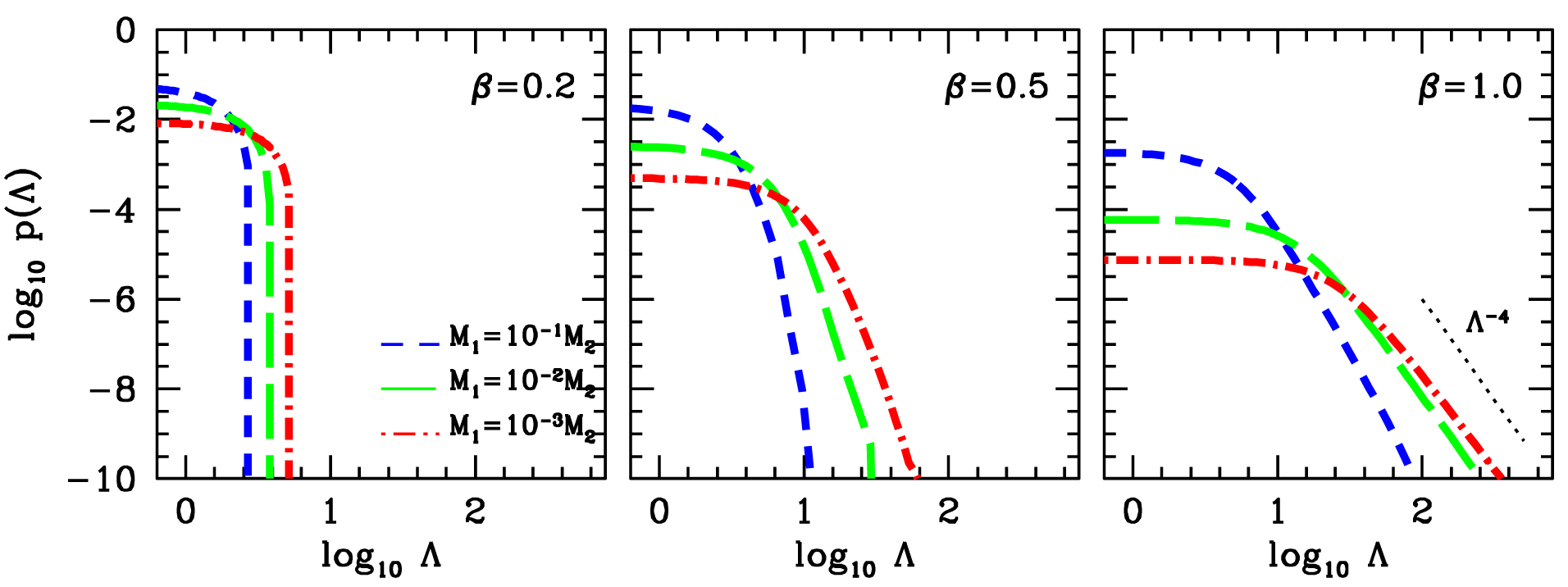}
\end{center}
\caption{Tidal force distribution $p({\bb \Lambda})$ arising from a population of extended substructures with a power-law mass function~(\ref{eq:pl}) with index $\alpha=-2$ and a normalization $B_0=1$. As expected from Equation~(\ref{eq:lamvar_mf}), populations with a size distribution with a power-law index $\beta>1/3$  exhibit a large-force behaviour that approaches asymptotically the point-mass tail $p({\bb \Lambda})\sim \Lambda^{-4}$ as $M_1\to 0$. In this regime the fluctuations of the force derivative can in principle reach arbitrarily-large values.}
\label{fig:pl_pl}
\end{figure*}
As highlighted in Section~\ref{sec:massfun}, understanding the dependence between the $p(\bb \Lambda$) and the substructure size is particularly interesting in cosmology. In the rarefied regime, where $\chi=q/\lambda_0=\pi^2 c^3 n/3=(\pi/6)(c/D)^3\ll 1$, it is possible to derive analytical expressions for the mean and the variance of the distribution using Equation~(\ref{eq:plam_hern_1}). After some algebra we find
\begin{eqnarray}\label{eq:lamav}
  \langle \Lambda\rangle &=& \int_0^{\lambda_0}\d^3\Lambda\,p_1({\bb \Lambda})\Lambda\\ \nonumber
  &=&\frac{4C}{\pi}q\int_0^{1/\chi}\d\xi\,\frac{\xi^3}{(1+\xi^2)^2}\big[1-\chi^{1/3}\xi^{1/3}\big]^2\\ \nonumber
   &= & C q \big[\frac{4}{\pi}(-5-\ln\chi)+ \frac{28}{3}\chi^{1/3}-\frac{16}{3\sqrt{3}}\chi^{2/3} +\mathcal{O}(\chi^{4/3})\big],
\end{eqnarray}
and
\begin{eqnarray}\label{eq:lamvar}
  \langle \Lambda^2\rangle &=& \int_0^{\lambda_0}\d^3\Lambda\,p_1({\bb \Lambda})\Lambda^2\\ \nonumber
  &=&\frac{4C}{\pi}q^2\int_0^{1/\chi}\d\xi\,\frac{\xi^4}{(1+\xi^2)^2}\big[1-\chi^{1/3}\xi^{1/3}\big]^2\\ \nonumber
  &= & C q^2 \big[\frac{2}{5\pi \chi}-3 +\frac{40}{3\sqrt{3}}\chi^{1/3}+\frac{8}{\pi}\chi-\frac{22}{3}\chi^{2/3} + \mathcal{O}(\chi^{4/3})\big].
\end{eqnarray}
Thus, in the limit $\chi\to 0$ one has $C\to 1$ and
\begin{eqnarray}\label{eq:lamvar2}
  \langle \Lambda\rangle &\simeq& \frac{4 q}{\pi}\ln(1/\chi)=\frac{8\pi}{3}\ln\big(\frac{3}{\pi^2c^3n}\big)GMn,\\\nonumber
  \langle \Lambda^2\rangle &\simeq& \frac{2 q^2}{5\pi\chi}=\frac{8\pi}{15}\frac{(GM)^2n}{c^3}.
\end{eqnarray}
Hence, the averaged tidal force has a mild logarithmic divergence as the substructure size decreases, whereas the magnitude of the tidal fluctuations strongly diverges in the limit $c\to 0$. Comparison of Equations~(\ref{eq:lamvar2}) and~(\ref{eq:F2_hern}) reveals that, while the variance of the combined force goes as $\langle F^2\rangle \sim c^{-1}$, the variance associated with the {\it derivative} of the force scales as $\langle \Lambda^2\rangle \sim c^{-3}$, suggesting that the smallest substructures may represent an important source of tidal heating.

\subsection{Substructures with power-law mass \& size functions}\label{sec:hern_mf_l}
Following Section~\ref{sec:hern_mf}, consider an ensemble of extended substructures with power-law mass~(\ref{eq:pl}) and size~(\ref{eq:cm}) functions, such that
$$n\to \int \int \frac{\d^2 n}{\d M\d c}\d M\d c,$$
with
\begin{eqnarray}\label{eq:nmc2}
\frac{\d^2 n}{\d M\d c}=B_0\bigg(\frac{M}{M_0}\bigg)^\alpha\delta\bigg[c-c_0\bigg(\frac{M}{M_0}\bigg)^\beta\bigg].
\end{eqnarray}
As a first step it is useful to calculate the function $\phi({\bf k})$ by integrating Equation~(\ref{eq:phih_l}) over the mass range $(M_1,M_2)$, which yields
\begin{eqnarray}\label{eq:phih_l_massfun}
  \phi({\bf k})&=&\frac{2\pi^2}{3}\frac{G B_0}{M_0^\alpha}k\int_{M_1}^{M_2}\d M\,M^{\alpha+1}\sum_{i=1}^3 J_i(k,M) \\ \nonumber
  &\equiv& Q'(k) k.
\end{eqnarray}
In the limit $c\to 0$ this function approaches asymptotically the point-mass behaviour, where $\sum_{i} J_i\to 1$ and 
\begin{eqnarray}\label{eq:q_massfun}
 \lim_{c\to 0}Q'(k)= q'\equiv \frac{2\pi^2}{3}\frac{G B_0}{M_0^\alpha}\times
\begin{cases}
\frac{M_2^{\alpha+2}-M_1^{\alpha+2}}{ \alpha+2}  & ,\alpha\ne - 2 \\ 
\ln\big( {M_2/M_1}\big)  & , \alpha=-2.
\end{cases}
\end{eqnarray}
Hence, the tidal force distribution induced by a population of point-mass particles with a broad mass spectrum ($M_2\gg M_1$) and a fixed normalization ($B_0={\rm const}.$) is dominated by the most massive particles in the ensemble if $\alpha>-2$, whereas for $\alpha<-2$ the least-massive ones dominate the combined tidal field. 

As shown in Section~\ref{sec:hern_mf}, in the case of extended substructures the size function, $c(M)$, determines the large-force behaviour of $p({\bb \Lambda})$. In a rarefied clumpy medium, where the separation between substructures is much larger than their individual sizes, $c\ll D$, one can use Equation~(\ref{eq:lamvar2}) to derive an analytical expression of the variance 
\begin{eqnarray}\label{eq:lamvar_mf}
  \langle \Lambda^2\rangle &\simeq& \frac{8\pi}{15}\frac{G^2B_0}{M_0^\alpha}\int_{M_1}^{M_2}\d M\, \frac{M^{\alpha+2}}{c(M)^3} \\ \nonumber
  &=&\frac{8\pi}{15}M_0^{3\beta-\alpha}\frac{G^2B_0}{c_0^3 }\times
  \begin{cases}
\frac{M_2^{3+\alpha-3\beta}-M_1^{3+\alpha-3\beta}}{3+\alpha -3\beta}  &,~ 3+\alpha-3\beta\ne 0 \\ 
\ln\big( {M_2/M_1}\big)  &,~ 3+\alpha-3\beta=0.
\end{cases}
\end{eqnarray}
Recall that the value of $\langle \Lambda^2\rangle$ provides a faithful representation of the magnitude of the tidal fluctuations. Crucially, Equation~(\ref{eq:lamvar_mf}) reveals two well-defined regimes of tidal perturbations
\begin{itemize}
\item {\bf Macrostructure-dominated tidal field} corresponds to substructure populations with a size function index $\beta < 1+\alpha/3$, such that $M_2^{3+\alpha-3\beta}-M_1^{3+\alpha-3\beta}\simeq M_2^{3+\alpha-3\beta}$, yielding a variance $\langle \Lambda^2\rangle\sim M_2^{3+\alpha-3\beta}$ which is dominated by the largest objects in the ensemble.
\item {\bf Microstructure-dominated tidal field}. In this case the least-massive substructures govern the large-force behaviour of $p({\bb \Lambda})$. This regime arises when the size function of substructures is sufficiently steep, $\beta \ge 1+\alpha/3$, which results in a force distribution with a variance that scales as $\langle \Lambda^2\rangle\sim M_1^{3+\alpha-3\beta}$, thus diverging in the limit $M_1\to 0$. Recall that a divergent variance means that the fluctuations of the tidal force can in principle reach arbitrarily-large values.
\end{itemize}

To illustrate this result we plot in Fig.~\ref{fig:pl_pl} the tidal force distribution associated with the CDM mass function outlined in Section~\ref{sec:cosmo_pf}. To derive $p({\bb \Lambda})$ we insert~(\ref{eq:phih_l_massfun}) into~(\ref{eq:pL_hern}) and integrate over $\d k$ numerically.
In these models the power-law index of the mass function~(\ref{eq:pl}) is $\alpha= -2$, which sets the transition between micro- and macro-dominated tidal force distributions at $\beta=1/3$. In addition, we choose $B_0=c_0=M_2=G=1$ for convenience. The remaining free parameter corresponds to the lower limit of the mass function, $M_1$, which determines the total number of substructures in the ensemble. Left, middle and right panels of Fig.~\ref{fig:pl_pl} show models with $\beta=0.2, 0.5$ and $1.0$, respectively. As expected from Equation~(\ref{eq:lamvar_mf}), substructure populations with a shallow size function $\beta=0.2<1/3$ exhibit distributions $p({\bb \Lambda})$ truncated at large forces. These models are macro-structure dominated, e.g. lowering the value of $M_1$ by two orders of magnitude merely increases the maximum tidal force by a factor $\sim 2$. In contrast, substructure ensembles in the middle and right panels have size functions with $\beta>1/3$ and are therefore microstructure dominated. In this regime decreasing the value of $M_1$ leads to a sharp increase in the maximum tidal force generated by the substructure population as a whole. As expected, in the limit $M_1\to 0$ the large-force tail of the distribution approaches asymptotically the point-mass behaviour $p({\bb \Lambda})\sim \Lambda^{-4}$, which according to Equation~(\ref{eq:lamvar2}) leads to a fluctuation spectrum with a divergent variance.

The transition between macro- and micro-structure dominated tidal forces in the $\Lambda$CDM paradigm, where $\alpha\approx -1.9$ (e.g. Springel et al. 2008), corresponds to $\beta\simeq 1+\alpha/3 \approx 0.37$. The relatively steep ($\beta\simeq 0.5$) size function of subhaloes found in the Via Lactea II dark matter-only simulations (Erkal et al. 2016; Diemand et al. 2008) indicates that {\it the largest fluctuations of the tidal field of a galaxy will be caused by the smallest \& most-abundant subhaloes}. Given that ultra-high-resolution $\Lambda$CDM simulations of structure formation predict subshaloes with masses as low as $M_1\sim 10^{-6}M_\odot$ (Ishiyama et al. 2010 and references therein), we expect microhaloes to induce a significant tidal heating of self-gravitating objects moving across galaxies like the Milky Way, where the upper limit of the subhalo mass function is $M_2\sim 10^{11}M_\odot\sim 10^{17}M_1$, an issue that we study in more detail below.

\begin{figure*}
\begin{center}
\includegraphics[width=162mm]{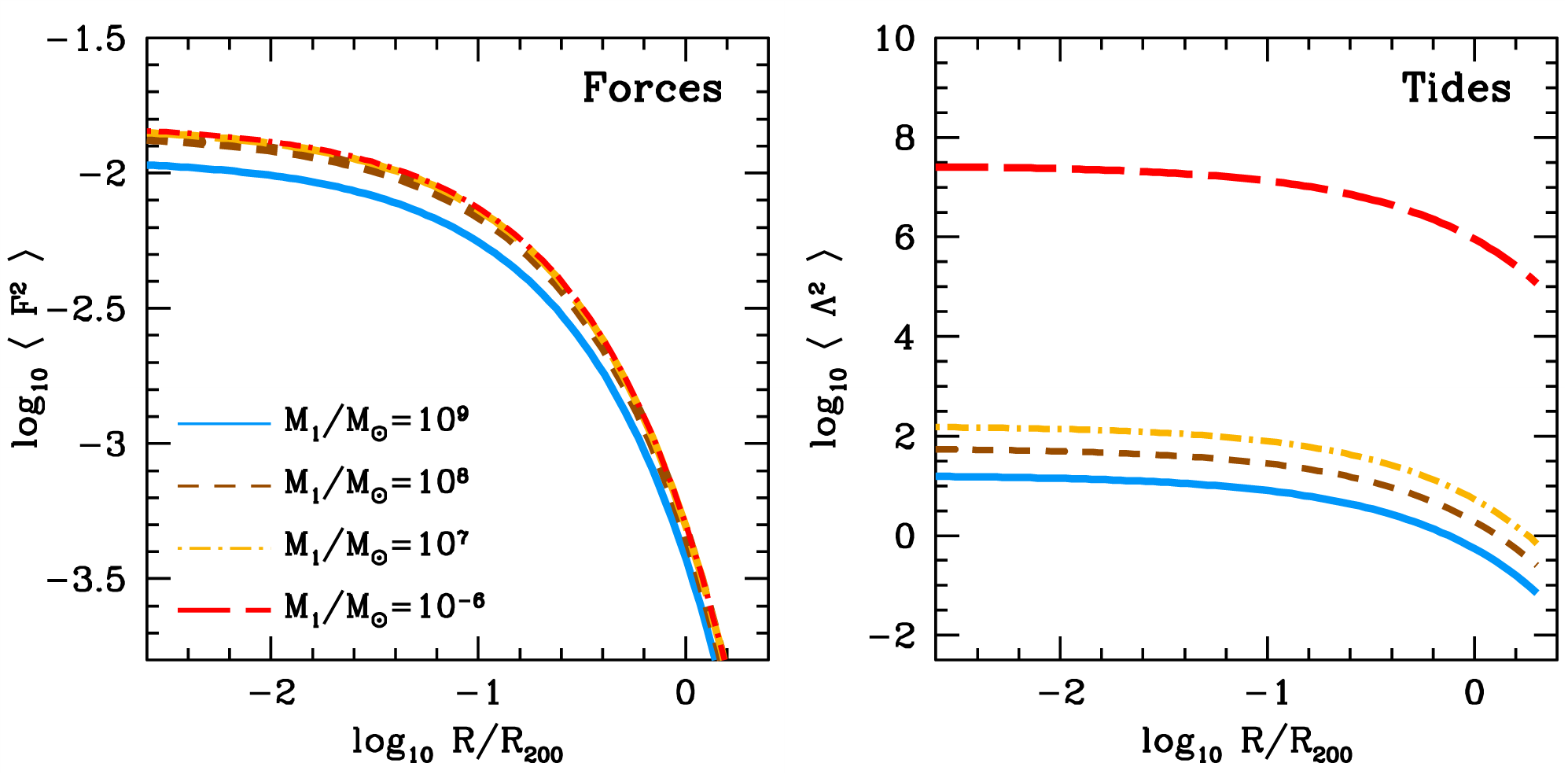}
\end{center}
\caption{Variance of the combined force (left panel) and tidal (right panel) distributions derived from Equations~(\ref{eq:F2_cosmo}) and~(\ref{eq:L2_cosmo}), respectively, for Aquarius subhaloes in different mass bins. 
  The upper limit of the mass function is fixed to $M_2/M_{\odot}=10^{10}$, where as the low-mass limit is $M_1/M_\odot=10^{9}, 10^{8}, 10^{7}$, and $10^{-6}$. Force $({\bf F}$) and tidal ($\bb \Lambda$) vectors are measured in units of $GM_{200}/R_{200}^2$ and $GM_{200}/R_{200}^3$, respectively. Note that the combined forces are dominated by the most massive objects of the subhalo population, while tidal forces are extremely sensitive to the low-end of the mass function. In particular, extrapolating of the Aquarius mass function down to planet-mass scales $M\sim 10^{-6}M_\odot$(red-dashed lines) yields tidal fluctuations $\sim 10^6$ times larger than that generated by satellites with $M\gtrsim 10^9M_\odot$. In contrast, the variance of the force distribution $\langle F^2\rangle$ is insensitive to subhaloes with $M\lesssim 10^7 M_\odot$ (dotted-dashed lines). }
\label{fig:aver}
\end{figure*}

\section{Discussion}\label{sec:dis}
One of the key predictions from $\Lambda$CDM simulations is the existence of myriads of self-gravitating DM clumps devoid of baryons (i.e. ``dark'').  
The mathematical framework outlined in Section~\ref{sec:massfun} provides a flexible tool to study the fluctuations in the gravitational field induced by a large population of these objects, allowing us to explore scales that are far beyond those reached by self-consistent cosmological $N$-body models of Milky Way-like galaxies.

\subsection{CDM microhaloes}\label{sec:micro}
As an illustration let us construct substructure ensembles that mimic the subhalo population found in the Aquarius simulations (Springel et al. 2008). The number density profile is well described by an Einasto profile
\begin{eqnarray}
  n(R)=n_0~\exp\bigg\{-\frac{2}{\gamma}\bigg[ \bigg(\frac{R}{R_{-2}}\bigg)^\gamma-1\bigg]\bigg\},
  \label{eq:nr}
  \end{eqnarray}
where $n_0$ is an overall constant that scales with the mass range considered, $\gamma=0.678$, $R_{-2}=0.81R_{200}=199\kpc$, and $R_{200}=246\kpc$. The virial mass of the Aquarius halo at redshift $z=0$ is $M_{200}=1.84\times 10^{12}M_\odot$. 

The mass function of these objects follows a power-law
\begin{eqnarray}
  \frac{\d N}{\d M}(M)=a_0\bigg(\frac{M}{M_0}\bigg)^\alpha,
  \label{eq:dndm}
  \end{eqnarray}
with $a_0=3.26\times 10^{-5}M_\odot^{-1}$, $M_{0}=2.52\times 10^{7}M_\odot$ and $\alpha=-1.9$ (Springel et al. 2008). Combination of Equations~(\ref{eq:nr}) and~(\ref{eq:dndm}) yields
\begin{eqnarray}
  \frac{\d n}{\d M}(R,M)=B_0\bigg(\frac{M}{M_0}\bigg)^\alpha\exp\bigg\{-\frac{2}{\gamma}\bigg[ \bigg(\frac{R}{R_{-2}}\bigg)^\gamma-1\bigg]\bigg\},
  \label{eq:nrm}
  \end{eqnarray}
where $B_0=2.02\times 10^{-13}M_\odot^{-1}\kpc^{-3}$ (see also Erkal et al. 2016; Han et al. 2016). Extrapolation of Equation~(\ref{eq:nrm}) down to the planet-mass scales of CDM microhaloes predicts an enormous number of low-mass substructures within the host virial radius
\begin{eqnarray}
  N=4\pi \int_0^{R_{200}}R^2 \d R\,\int_{M_1}^{M_{200}}\d M\, \frac{\d n}{\d M}\sim 10^{15} ~~~~{\rm for}~~M_1=10^{-6}M_\odot.
  \label{eq:ntot}
\end{eqnarray}

The number of subhaloes drastically decreases at the high end of the halo mass function. Given that the statistical models outlined in Section~\ref{sec:fluctu} assume $N\gg 1$, our analysis is bound to fail when the minimum mass becomes comparable to the host mass. Issues introduced by low-number statistics can be largely removed by choosing an upper-mass limit $M_2\lesssim 10^{-3}M_{200}$, which yields $N\gtrsim 100$ for $M_2\gg M_1$.

Following Erkal et al. (2016) let us adopt a power-law size function~(\ref{eq:cm}) with $c_0=0.53\kpc$ and $\beta=0.5$, which matches the relation between bound mass and scale radius of the subhaloes found in the Via Lactea II simulation (Diemand et al. 2008). The average size of subhaloes in a mass bin $M\in (M_1,M_2)$ is
\begin{eqnarray}
  \langle c \rangle=\frac{\int_{M_1}^{M_2}\d M\, c(M)\frac{\d n}{ \d M}}{\int_{M_1}^{M_2}\d M\, \frac{\d n}{\d M }} =\frac{c_0}{M_0^\beta}\frac{1+\alpha}{1+\alpha+\beta}\bigg(\frac{M_2^{1+\alpha+\beta}-M_1^{1+\alpha+\beta}}{M_2^{1+\alpha}-M_1^{1+\alpha}}\bigg),
  \label{eq:cma}
\end{eqnarray}
which for $M_2/M_\odot=10^{10}$ and $M_1/M_\odot=10^{9},10^{8}, 10^7$ and $10^{-6}$ yields $\log_{10}\langle c\rangle/\kpc=0.69, 0.29, -0.17$ and -6.64, respectively. It is worth noting that extrapolation of the power-law relation~(\ref{eq:cm}) to planet-sized masses is grossly consistent with the size of microhaloes expected from an inflation-produced primeval fluctuation spectrum (Berezinsky et al. 2003).

Fig.~\ref{fig:aver} shows the variance of the force $\langle F^2\rangle$ (left panel) and tidal $\langle \Lambda^2\rangle$ (right panel) distributions induced by subhaloes in the above mass bins. Under the local approximation (see Appendix A) the variance of the force distribution follows from Equations~(\ref{eq:F2_pl}) and~(\ref{eq:nrm}) by setting $\alpha=-1.9$ and $\beta=0.5$, which yields
\begin{eqnarray}
  \langle F^2 \rangle(R) \approx 6.981 \frac{G^2 M_0^{2.4}B_0}{c_0}\frac{n(R)}{n_0}(M_2^{0.6}-M_1^{0.6}).
  \label{eq:F2_cosmo}
\end{eqnarray}
Note that the variance of the force distribution converges asymptotically towards $\langle F^2 \rangle\sim M_2^{0.6}$ for $M_2\gg M_1$. As a result, Fig.~\ref{fig:aver} shows that low-mass ($M\lesssim 10^7 M_\odot$) subhaloes contribute negligibly to the stochastic fluctuations of the force field.
This has important implications for the detection of micro-haloes using experiments that rely on the scattering of tracer orbits (see Introduction).
Note also that $\langle F^2 \rangle$ reaches its maximum at the host centre. However, this result must be taken with caution as we show in Appendix A that at small galactocentric distances the average separation between massive subhaloes is larger than the distance on which $n(R)$ varies significantly, $D\gtrsim d=|\nabla n/n|^{-1}$, which invalidates the locality assumption.

In contrast, the Aquarius tidal field is remarkably sensitive to the low-mass end of the subhalo mass function. Indeed, combination of Equations~(\ref{eq:lamvar_mf}) and~(\ref{eq:nrm}) yields
\begin{eqnarray}
  \langle \Lambda^2 \rangle(R) \approx 4.189 \frac{G^2 M_0^{3.4}B_0}{c_0^3}\frac{n(R)}{n_0}\bigg(\frac{1}{M_1^{0.4}}-\frac{1}{M_2^{0.4}}\bigg),
  \label{eq:L2_cosmo}
\end{eqnarray}
which diverges as $\langle \Lambda^2 \rangle \sim M_1^{-0.4}$ in the limit $M_1\to 0$. Thus, tidal field fluctuations are completely dominated by the smallest, most-abundant subhaloes in the ensemble. As shown in the right panel of Fig.~\ref{fig:aver}, extrapolating the subhalo mass function from $M\sim 10^7M_\odot$ down to the planet-scales of microhaloes increases the variance of the tidal field by $\sim 6$ orders of magnitude, suggesting that the existence of `dark' satellites could in principle be tested with observational experiments that measure fluctuations in the Galactic tidal field (see \S\ref{sec:obs} below).

\subsection{Alternative dark matter-particle models}\label{sec:dm}
The mass and size functions of the subhalo population are determined by the quantum attributes of the DM particle model. In this Section we provide a simplistic estimate of the spectrum of gravitational perturbations that one would expect to measure in Milky Way-sized haloes made of warm particles (\S\ref{sec:WDM}), or ultra-light scalar fields (\S\ref{sec:WDM}). Section~\ref{sec:comp} provides a brief comparison of stochastic field properties in CDM, WDM and wave-DM haloes.

\subsubsection{Warm Dark Matter (WDM) subhaloes}\label{sec:WDM}
Cosmological simulations of structure formation show a strong drop in the number of subhaloes with masses $M\lesssim M_1$, where $M_1$ is the Jeans mass scale associated with the free-streaming length of the DM particle model. For CDM haloes made of neutralinos, the lightest stable supersymmetric particle ($m_{\rm CDM}\sim 100$ GeV, Jungman et al. 1996), the mass cut-off is expected at $M_1\sim 10^{-6}M_\odot$ (e.g. Schneider et al. 2013).
In contrast, WDM particles decouple later and have non-negligible thermal velocities, which leads to a lower cut-off of the mass function that increases inversely with the particle mass. The COCO project (Bose et al. 2016) compares the properties of small-scale structures in CDM and WDM $N$-body simulations that incorporate current Lyman-$\alpha$ forest constraints ($m_{\rm WDM}\ge 3.3\kev$; Viel et al. 2013), finding a lower mass cut-off at $M_1\sim 3\times 10^8 M_\odot$.

The suppression of structure formation at low masses leads to a relatively shallow subhalo mass function (Schneider et al. 2012; Angulo et al. 2013; Lovell et al. 2014)
\begin{eqnarray}\label{eq:nwdm}
  \frac{\d n_{\rm WDM}}{\d M}=\frac{\d n_{\rm CDM}}{\d M}\bigg(1+\frac{\mu M_1}{M}\bigg)^\eta,
\end{eqnarray}
with $\mu\simeq 2.2$ and $\eta\simeq -0.75$ (Ludlow {\it personal comm.}). Hence, the slope of the function~(\ref{eq:nwdm}) rolls from the CDM power-law index $\alpha_{\rm WDM}\approx \alpha_{\rm CDM}= -1.9$ at large masses $M\gg \mu M_1$, down to $\alpha_{\rm WDM}\approx \alpha_{\rm CDM}-\eta\simeq -1.2$ at $M\sim \mu M_1$.

In addition, numerical simulations of structure formation in the WDM paradigm show that (i) the internal density profile of cold and warm dark matter subhaloes can be well fitted by a Hernquist (1990) model (Macci\`o et al. 2012), (ii)
WDM subhaloes systematically have lower central densities and larger scale radii relative to their CDM counterparts of the same mass (Lovell et al. 2014), and (iii) the fractional change of scale radii between WDM and CDM models barely depends on subhalo mass (e.g. see Fig. 9 Bose et al. 2016). Thus, one can incorporate WDM subhalo properties in the size function~(\ref{eq:cm}) by setting $c_{0,{\rm WDM}}\gtrsim c_{0,{\rm CDM}}$, while keeping the logarithmic slope fixed $\beta_{\rm WDM}\sim\beta_{\rm CDM}=0.5$.

It is straightforward to show that deviations of the mass function~(\ref{eq:nwdm}) from a scale-free relation~(\ref{eq:dndm}) have a minor impact on the variance of the force and tidal distributions. Replacing the mass function~(\ref{eq:dndm}) by~(\ref{eq:nwdm}) and following the same steps as in \S\ref{sec:micro} yields
\begin{eqnarray}\label{eq:fvar_wdm}
  \frac{\langle F^2\rangle_{\rm WDM}}{\langle F^2\rangle_{\rm CDM}}&=&\frac{\int_{M_1}^{M_2}\d M\, M^{\alpha+2-\beta}\big(1+\frac{\mu M_1}{M}\big)^\eta}{\int_{M_1}^{M_2}\d M\, M^{\alpha+2-\beta}}\frac{c_{0,{\rm CDM}}}{c_{0,{\rm WDM}}} \\ \nonumber
  &\simeq& \frac{c_{0,{\rm CDM}}}{c_{0,{\rm WDM}}},
  \end{eqnarray}
and
\begin{eqnarray}\label{eq:lamvar_wdm}
  \frac{ \langle \Lambda^2\rangle_{\rm WDM}}{\langle \Lambda^2\rangle_{\rm CDM}}&=&\frac{\int_{M_1}^{M_2}\d M\, M^{\alpha+2-3\beta}\big(1+\frac{\mu M_1}{M}\big)^\eta}{\int_{M_1}^{M_2}\d M\, M^{\alpha+2-3\beta}}\bigg(\frac{c_{0,{\rm CDM}}}{c_{0,{\rm WDM}}}\bigg)^3\\ \nonumber
 & \simeq& 0.75\bigg(\frac{c_{0,{\rm CDM}}}{c_{0,{\rm WDM}}}\bigg)^3,
  \end{eqnarray}
where we have adopted $\alpha=-1.9$, $\beta=0.5$, $\mu=2.2$ and $\eta=-0.75$, within a fixed mass range $M_1\ll M_2$.

Given that the relative difference in size has order unity $c_{0,{\rm WDM}}/c_{0,{\rm CDM}}\sim 1$--2 (Lovell et al. 2014; Bose et al. 2016), 
 the above estimates indicate that the dominant aspect that distinguishes CDM and WDM haloes corresponds to the lower cut-off of the mass function, $M_{1,{\rm CDM}}\lll M_{1,{\rm WDM}}$, which may lead to significant differences in the amplitude of tidal fluctuations (see right panel of Fig.~\ref{fig:aver}).

\subsubsection{Ultra-light axion (ULA) subhaloes}\label{sec:ULA}
Ultra-light scalars with masses in the range $10^{-24} \le m_a/\ev \le 10^{-20}$ have been proposed as viable DM candidates (e.g. Hu et al. 2000; Marsh 2016 and references therein). As in WDM models, ULAs also suppress linear power on scales below the mass scale associated with the particle free-streaming length (e.g. Marsh 2016), which leads to a shallow mass function at low masses $M\lesssim M_1$, approaching the CDM scale-free relation~(\ref{eq:dndm}) at $M\gtrsim M_1$ (Schive \& Chiueh 2017).

An important aspect of ULA models is that, while CDM and WDM models are governed by the collision-less Boltzmann equation, self-gravitating ULA haloes follow the Schr\"{o}dinger-Poisson equation, thus behaving as a single coherent wave function on scales below the de Broglie wavelength (Hu et al. 2000). In particular, at radii $r\lesssim r_c$ ULA haloes form a class of pseudo-soliton known as an oscillaton, or `axion star' (Ruffini \& Bonazzola 1969; Guzm\'an \& Ure\~na-L\'opez 2004). At larger radii $r\gtrsim r_c$ the density profile of ULA subhaloes becomes similar to that found in CDM haloes (Schieve et al. 2014a,b; Schwabe et al. 2016).
Numerical simulations show that the soliton is orders of magnitude denser than the surrounding, CDM-like halo envelope, and that the size of the soliton and the halo mass are anti-correlated (Schive et al. 2014a,b; Du et al. 2017)
\begin{eqnarray}\label{eq:rc}
  r_c=1.6\kpc \bigg(\frac{m_a}{10^{-22}\ev}\bigg)^{-1}\bigg(\frac{M}{10^8M_\odot}\bigg)^{-1/3}.
  \end{eqnarray}
Therefore, for values $m_a\lesssim 10^{-22}\ev$ the predicted size function has a negative power-law slope $\beta=-1/3$, whereas for $m_a\gg 10^{-22}\ev$ axion models converge towards the size \& mass functions of CDM subhaloes (see Gonz\'ales-Morales et al. 2016 for a compilation of current observational bounds on $m_a$).

\begin{figure}
\begin{center}
\includegraphics[width=84mm]{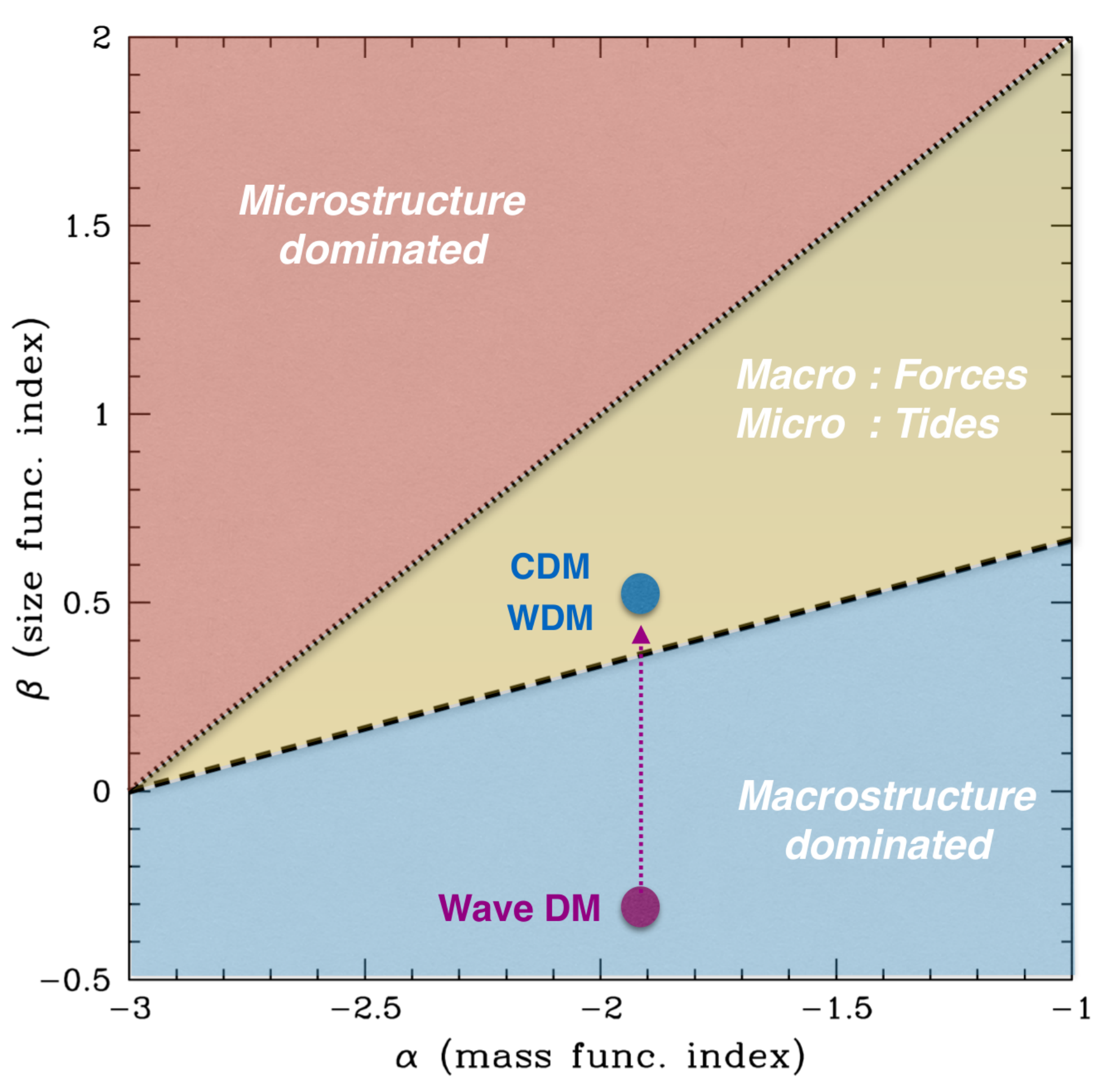}
\end{center}
\caption{Combination of power-law indices of the mass ($\alpha$) and size ($\beta$) functions that determine the large-force behaviour of the force and tidal fields induced by a large population of substructures in Milky Way-like haloes. Dashed and dotted lines represent the transition between macro- and microstructure-dominated force ($3+\alpha+\beta=0$) and tidal ($3+\alpha+3\beta=0$) distributions, $p({\bf F}$) and $p(\bb \Lambda)$, respectively. For ease of reference we mark the values found in numerical simulations of structure formation within the $\Lambda$CDM paradigm (blue circle, Springel et al. 2008; Diemand et al. 2007), Wave DM (purple circle, Schive et al. 2014a) and Warm DM (Lovell et al. 2014; Bose et al. 2016).  Wave-like models converge towards the CDM limit as the particle mass $m_a\gg 10^{-22}\ev$ (dotted arrow).}
\label{fig:ab}
\end{figure}

\subsubsection{Gravitational fluctuations of CDM,WDM \& ULA haloes}\label{sec:comp}
Fig.~\ref{fig:ab} plots a schematic diagram that summarizes the general behaviour of the force and tidal distributions for subhalo populations extracted from CDM, WDM and ULA simulations of Milky Way-like haloes. For simplicity, we adopt power-law mass and size functions with indices $\alpha$ and $\beta$, respectively. In addition, we assume that Equation~(\ref{eq:fh}) provides a reasonable representation of the force induced by individual subhaloes. This approximation is particularly poor for ULA subhaloes, which contain two relevant scales, the soliton radius ($r_c$) and the size of the outer halo envelope (Schive et al. 2014a). In our estimates we set the subhalo scale-radius $c=r_c$, where $r_c$ is given by Equation~(\ref{eq:rc}), a choice that becomes less accurate as the ULA models approach the CDM limit $r_c\to 0$ ($m_a\to \infty$).

As shown in previous Sections, whether the fluctuations of the force \& tidal fields are dominated by the most massive satellites (blue), or in contrast by the smallest and most abundant subhaloes (red), mainly depends one the combination of the power-law indices $\alpha$ and $\beta$. Interestingly, CDM and WDM models are located in an intermediate regime (yellow area), where the fluctuations of the gravitational field $p({\bf F})$ are governed by the largest satellites with $M\sim M_{200}$, and the spectrum of tidal forces $p({\bb \Lambda})$ by subhaloes with masses $M\sim M_1$, where $M_1$ is the lower cut-off end of the subhalo mass function. The fact that the soliton size is anticorrelated with the halo mass ($\beta<0$) solidly places ULA models with $m_a\lesssim 10^{-22}\ev$ in the macroscopic-dominated region of the diagram. For axion masses $m_a\gg 10^{-22}\ev$ ULA models are expected to approach the CDM fluctuation regime, which we mark with a dotted arrow for illustration.

The results encapsulated in Figs.~\ref{fig:aver} and~\ref{fig:ab} suggest that probing the Galactic tidal field may offer promising avenues to test the particle-nature of DM. In particular, detecting tidal fluctuations as strong as suggested by the right panel of Fig.~\ref{fig:aver} would favour the existence of a very large number of dense microhaloes, thus simultaneously falsifying both WDM and ULA models.
In contrast, a scenario where the spectrum of tidal forces can be accounted by the most massive (likely visible) satellites would be a clear indication of a macroscopic-dominated subhalo population, lending support to ULA predictions.
Finally, an intermediate scenario where the tidal field shows weak, but non-negligible fluctuations beyond those generated by the visible satellite population would put constraints on the low-end of the WDM subhalo mass function and rule out ULA models.

\subsection{Observational tests with wide binaries}\label{sec:obs}
Due to their low binding energies wide stellar binaries can be easily disrupted by the tidal field of the host galaxy (Heggie 1975; Bahcall et al. 1985; Chanam\'e \& Gould 2004). Here we use the Aquarius subhalo population described in \S\ref{sec:micro} to illustrate to what extent wide binaries may be used to constrain the low-end of the subhalo mass function.

In a reduced-mass frame Equation~(\ref{eq:eqmots}) 
  describes the separation ${\bf R}'={\bf R}_1-{\bf R}_2$ of
  two stars in a binary system with masses $m_1$ and $m_2$, respectively (see \S2 of Pe\~narrubia et al. 2016). The self-gravitating potential has a Keplerian form, $\Phi_s(R')=-Gm_b/R'$, where $m_b=m_1+m_2$ is the binary mass. Given that in Milky Way-size haloes $\langle \Lambda^2\rangle^{1/2}\gg GM_{200}/R_{200}^3$ (see right panel of Fig.~\ref{fig:aver}),  we can safely neglect the contribution of the smooth tidal field to the spectrum of tidal fluctuations.  
  The separation at which the binary's self-gravity becomes comparable to the amplitude of the external tidal fluctuations can be estimated from Equation~(\ref{eq:Flambda}) as $Gm_b/R'^2_{\rm max}=\langle \Lambda^2\rangle^{1/2} R'_{\rm max}$, which translates into a semi-major axis $2a_{\rm max}=R'_{\rm max}$. 
 Inserting the values of $B_0$, $c_0$ and $M_0$ found in the Aquarius and Via Lactea simulations into Equation~(\ref{eq:L2_cosmo}) and adopting a fiducial binary mass $m_b=1M_\odot$ yields
  \begin{eqnarray}\label{eq:rb}
    a_{\rm max}(R)\sim\frac{1}{2}\bigg(\frac{Gm_b}{\langle \Lambda^2\rangle^{1/2}}\bigg)^{1/3}\approx 0.9{\rm pc}\bigg[\frac{n(R)}{n_0}\bigg]^{-1/6}\bigg(\frac{M_1}{10^{-6}M_\odot}\bigg)^{1/15},
  \end{eqnarray}
  where $n(R)$ is the number density profile given by Equation~(\ref{eq:nr}) and $R$ is the galactocentric radius.

Binary stars with semi-major axes $a\gtrsim a_{\rm max}$ experience fluctuations of the tidal field with a magnitude comparable to the self-gravity of the system. 
For $M_1=10^{-6}M_\odot$ the separation at which one may expect to observe tidal perturbations varies from $a_{\rm max}(R=0)\simeq 0.6\pc$, up to $a_{\rm max}(R_{200})\simeq1\pc$ for binary stars located at the centre and the virial radius of the host halo, respectively. These values are comparable to the average separation between microhaloes, $1.1\lesssim D/\pc \lesssim 3.2$. In contrast, increasing the minimum subhalo mass up to $M_1=10^{7}M_\odot$ yields $4\lesssim a_{\rm max}/\pc\lesssim 7$, while the averaged distance between subhaloes grows up to kiloparsec-scales $9\lesssim D/\kpc \lesssim 23$.

Interestingly, Oelkers et al. (2017) and Oh et al. (2017) have recently identified  a large number of co-moving stellar pairs in the Tycho-Gaia Astrometric Solution (TGAS) (Gaia Collaboration et al. 2016) with separations $>1\pc$. 
The back-of-the-envelope estimate~(\ref{eq:rb}) suggests that the survivability of these systems may depend on the number density and lower cut-off mass of CDM microhaloes. This is an exciting possibility, which we plan to explore in a separate contribution with the aid of $N$-body models that follow the dynamical evolution of stellar pairs subject to a stochastic tidal field $p({\bb \Lambda})$. Such code may provide a useful tool to model the observed kinematics of Milky Way binaries within a Bayesian framework.

\section{Summary}\label{sec:sum}
Following the dynamical evolution of an arbitrarily-large number of self-gravitating systems is computationally unfeasible with current $N$-body methods. This shortcoming affects several fields of Astronomy and Cosmology which must typically deal with large populations of objects covering wide dynamical ranges.
Here we present a statistical technique for deriving the spectrum of random fluctuations of the combined force and tidal fields generated by statistical ensembles of extended substructures with known mass and size functions. Our work follows up the method originally devised by Holtsmark (1919) to study the motion of charged particles in an ionized plasma.

As a first application, we derive the force distribution induced by a large population of CDM subhaloes distributed across a Milky Way-like halo. Typically, the lowest-mass subhaloes that can be resolved in cosmological $N$-body simulations of Galaxy-sized haloes have masses $M\gtrsim 10^6 M_\odot$ (e.g. Diemand et al. 2007, Springel et al. 2008) and follow power-law size and mass functions.
Here we show that these objects generate force distributions, $p({\bf F})$, governed by the most massive satellites in the ensemble, and tidal distributions, $p({\bb \Lambda})$, that are completely dominated by the smallest and most abundant subhaloes.
Extrapolating the size and mass scaling relations $\sim 12$ orders of magnitude in mass, down to planet-mass scales of microhaloes ($M\sim 10^{-6} M_\odot$) indicates that $N$-body simulations that do not resolve the free-streaming scale of CDM particles greatly underestimate the magnitude of the tidal field fluctuations. This limitation may have a significant impact on dynamical processes that remain poorly known, such as star formation in molecular clouds, dissolution of open clusters in galactic tidal fields, tidal heating of thin discs, etc.

Crucially, our results open up the possibility to test the existence of planet-size microhaloes via experiments that are sensitive to tidal force fluctuations induced by dark matter substructures, e.g. by measuring the heating rate of weakly-bound gravitational systems in different regions of the stellar halo, as well as in the dwarf spheroidal galaxies of the Milky Way (see also Pe\~narrubia et al. 2010). This type of experiments may also constrain a wide range of cosmological theories where dark matter is made of `warm' particles or ultra-light axions (see \S\ref{sec:dm}).

The mathematical technique explored in this paper provides a computationally-efficient tool to model the gravitational perturbations induced by an arbitrarily-large number of CDM microhaloes. In follow-up contributions we plan to run numerical experiments that follow the trajectories of tracer particles in smooth galactic potentials that incorporate the force and tidal distributions derived in Sections~\ref{sec:fluctu} and~\ref{sec:tides} as external stochastic fields. Such codes may provide a useful tool to model observational data (e.g. the separation function of wide binaries, see \S\ref{sec:obs}) at a low computational cost.

Finally it is worth stressing that our statistical method can also be used to describe stochastic fluctuations in the force \& tidal fields generated by {\it baryonic} substructures covering a wide range of mass and sizes, such as Giant Molecular Clouds, MACHOs, black holes, stellar clusters, individual stars, planetoids, etc, which may have several applications in areas other than cosmology.

\section{Acknowledgements}
It is a pleasure to thank Denis Erkal, Doddy Marsh, Florent Renaud, Paco Prada and Mark Lovell for their comments \& suggestions, which have greatly improved this paper. Also many thanks to Aaron Ludlow for sharing the results of his analysis and the stimulating discussions with our colleagues in Newcastle.

{}
\appendix
\section{The local approximation}\label{sec:local}
In Sections~\ref{sec:fluctu} and~\ref{sec:tides} we derive the distributions $p({\bf F})$ and $p({\bb \Lambda})$, respectively, under the assumption that the number density of substructures remains constant across the entire system, and that $n=n({\bf R})$ corresponds to the value measured at the galactocentric location of the tracer particle, ${\bf R}$. Although the {\it local approximation} greatly simplifies analytical derivations\footnote{Section~4 of Kandrup (1980) provides solutions for force distribution of an inhomogeneous ensemble of point-masses where $n(r)\sim r^p$.} of the Fourier transform $\phi({\bf k})$, in general it fails on scales $r\gtrsim d$, where $d=|\nabla n/n|^{-1}$ is the distance associated with the spatial variation of the number density profile~(\ref{eq:nr})
\begin{eqnarray}
  d(R)\equiv \bigg|\frac{\nabla n}{n}\bigg|^{-1}_R=\frac{R_{-2}}{2}\bigg(\frac{R}{R_{-2}}\bigg)^{1-\gamma}.
  \label{eq:d}
  \end{eqnarray}
The fact that $\gamma<1$ in cosmological simulation of structure formation implies that the scale-length $d$ decreases towards the centre of the potential, suggesting that the local approximation may not be valid at small galactocentric radii.

\begin{figure}
\begin{center}
\includegraphics[width=84mm]{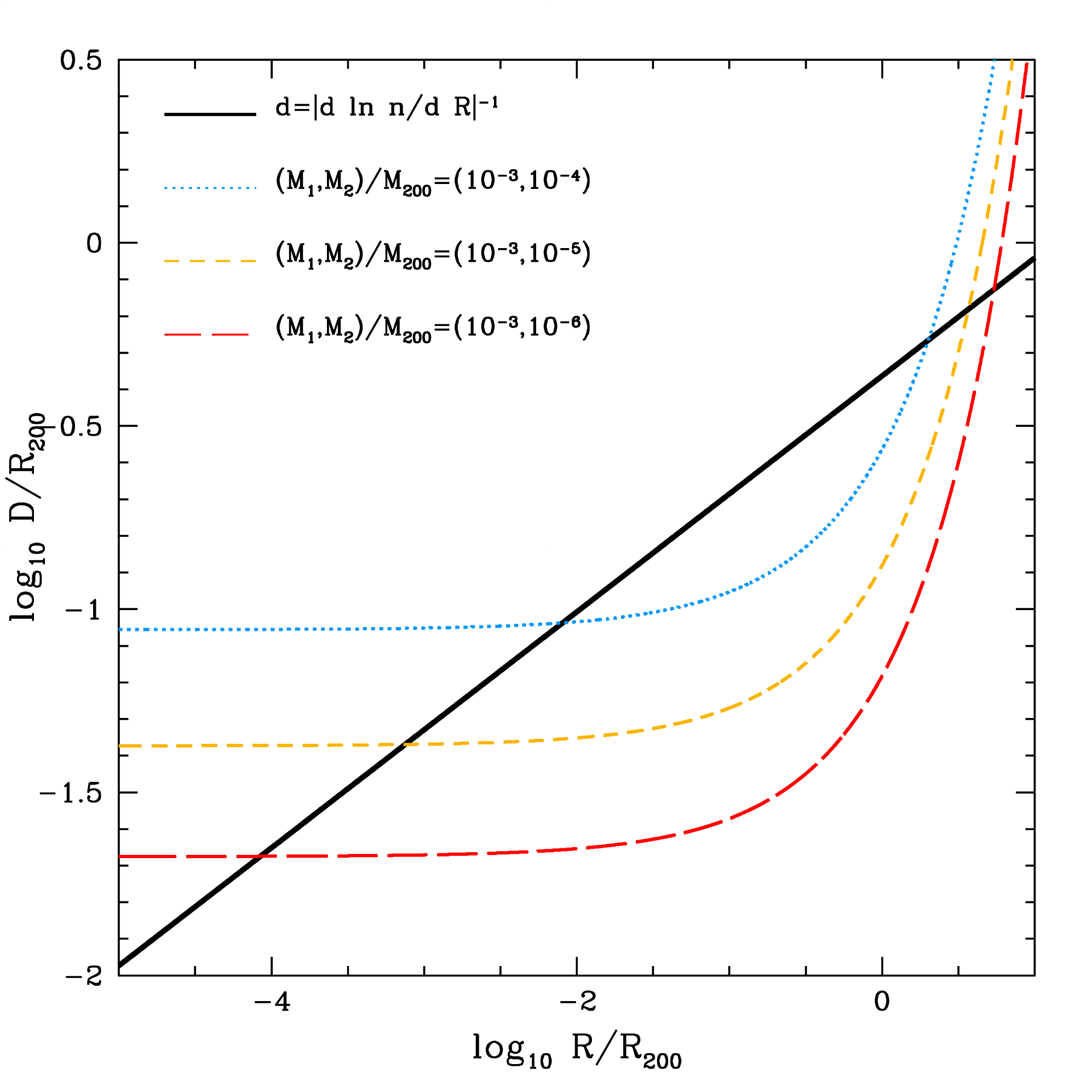}
\end{center}
\caption{Distance at which the probability to find the closest particle~(\ref{eq:pr_closest}) peaks, $D$, as a function of galactocentric distance for subhaloes within three different mass bins. The scale-length $d\equiv |\nabla n/n|^{-1}$ given by Equation~(\ref{eq:d}) is shown with a black solid line. Note that in a rarefied regime, where large-forces are dominated by the nearest substructure, the local approximation is valid on scales $d\gtrsim D$. For Aquarius subhaloes with $M<10^{-3}M_{200}$ this condition is satisfied within a radial range $10^{-2}\lesssim R/R_{200}\lesssim 1$. }
\label{fig:local}
\end{figure}

The second scale of relevance corresponds to the average separation between substructures, $D(R)$. In Section~\ref{sec:fluctu} we found that the large-force fluctuations generated by non-overlapping subhalo populations are dominated by the nearest substructure, which is typically located at a distance $D(R)\approx [2\pi n(R)]^{-1/3}$. Hence, the local approximation is justified for $d\gg D$, that is when the number density profile varies on scales that are much larger than the average separation between subhaloes.
Fig.~\ref{fig:local} shows a comparison between $d(R)$ and $D(R)$ measured from the Aquarius subhalo population as a function of galactocentric radius for three mass bins with $M_2/M_{200}=10^{-3}$ and $M_1/M_{200}=10^{-4},10^{-5}$ and $10^{-6}$.
As expected, at small radii the average distance between subhaloes (dashed and dotted lines) becomes larger than the scale-length $d$ (solid black line), which indicates that 
the local approximation fails at the inner-most regions of the galaxy. In addition, the sharp drop of $n(R)$ at $R\gtrsim R_{-2}\sim R_{200}$ leads to a fast growth of the average separation between subhaloes at large radii, such that $D\gtrsim d$ at $R\gtrsim R_{200}$.
Note that the radial range wherein the local approximation is valid increases as the low-mass limit of the mass function $M_1\to 0$ and the number of substructures in the sample grows. This is particularly true in the inner-most regions of the galaxy, where the number density profile~(\ref{eq:nr}) becomes flat. Fig.~\ref{fig:local} indicates that the local approximation may hold for objects with masses below $M\lesssim 10^{-3}M_{200}$ within a radial range $10^{-2}\lesssim R/R_{200}\lesssim 1$. Outside this range Equation~(\ref{eq:phi}) must be replaced by~(\ref{eq:phi_inhomog}), which accounts for the spatial variation of the number density profile at the location of the tracer particle (Kandrup 1980).

\end{document}